\documentclass[conference]{IEEEtran}
\usepackage{xspace}
\usepackage{subcaption}
\usepackage{listings}
\usepackage{xcolor}
\usepackage{url}
\usepackage[ruled,vlined,linesnumbered]{algorithm2e}
\usepackage{amsmath}
\usepackage[capitalise]{cleveref}
\SetKw{Break}{break}
\usepackage{booktabs}
\usepackage{pgf}
\usepackage[export]{adjustbox}
\usepackage{xparse}
\usepackage{multirow}
\usepackage[numbers,sort&compress]{natbib}
\newcommand{\Description}[1]{}

\ifCLASSINFOpdf
\else
\fi
\newif\ifdraft
\draftfalse

\ifdraft
  \newcommand{\todo}[1]{\textcolor{blue}{TODO: #1}}
  \newcommand{\note}[1]{\textcolor{red}{NOTE: #1}}
  \newcommand{\question}[1]{\textcolor{magenta}{QUESTION: #1}}
  \newcommand{\mkr}[1]{\textcolor{purple}{MKR: #1}}
  \newcommand{\lujo}[1]{\textcolor{blue}{Lujo: #1}}
  
  \newcommand{\anna}[1]{\textcolor{magenta}{Anna: #1}}
  \newcommand{\reviewfeedback}[1]{\textcolor{brown}{Review Feedback: #1}}
\else
  \newcommand{\todo}[1]{}
  \newcommand{\note}[1]{}
  \newcommand{\question}[1]{}
  \newcommand{\mkr}[1]{}
  \newcommand{\lujo}[1]{}
  
  \newcommand{\anna}[1]{}
  \newcommand{\reviewfeedback}[1]{}
\fi

\newcommand{\SysName}{MindReader\xspace}
\newcommand{\sysName}{MindReader\xspace}

\newcommand{\gramLength}{\ensuremath{n}\xspace}

\definecolor{codegreen}{rgb}{0,0.6,0}
\definecolor{codegray}{rgb}{0.5,0.5,0.5}
\definecolor{codeblue}{rgb}{0.1,0.4,0.82}
\definecolor{backcolour}{rgb}{0.95,0.95,0.95}

\lstdefinestyle{mystyle}{
  backgroundcolor=\color{backcolour},   
  commentstyle=\color{teal},
  keywordstyle=\color{purple},
  numberstyle=\tiny\color{codegray},
  stringstyle=\color{violet},
  basicstyle=\ttfamily\footnotesize,
  breakatwhitespace=false,         
  breaklines=true,                 
  captionpos=b,                    
  keepspaces=true,                 
  numbers=left,                    
  numbersep=5pt,                  
  showspaces=false,                
  showstringspaces=false,
  showtabs=false,                  
  tabsize=2
}

\Crefname{lstlisting}{Listing}{Listings}
\crefname{algorithm}{Algorithm}{Algorithms}
\Crefname{algorithm}{Algorithm}{Algorithms}

\NewDocumentCommand{\PassToPath}{}{\text{Pass2Path}\xspace}

\DeclareMathOperator{\Supp}{supp}
\NewDocumentCommand{\nats}{o o}%
  {\IfNoValueTF{#1}%
    {\ensuremath{\mathbb{N}}\xspace}%
    {\IfNoValueTF{#2}%
       {\ensuremath{[{#1}]}\xspace}%
       {\ensuremath{[{#1},{#2}]}\xspace}%
    }%
  }
\NewDocumentCommand{\emptyvec}{}{\ensuremath{\langle\rangle}\xspace}

\NewDocumentCommand{\getsr}{}{\ensuremath{\overset{\scriptscriptstyle\$}{\leftarrow}}\xspace}
\NewDocumentCommand{\strLen}{g}{\ensuremath{\mathopen{\lvert}{#1}\mathclose{\rvert}}\xspace}
\NewDocumentCommand{\vecLen}{g}{\ensuremath{\mathopen{\lvert}{#1}\mathclose{\rvert}}\xspace}
\NewDocumentCommand{\segsLen}{}{\ensuremath{\ell}\xspace}
\NewDocumentCommand{\segIdx}{o}{\ensuremath{i\IfNoValueF{#1}{_{#1}}}\xspace}
\NewDocumentCommand{\segIdxIdx}{}{\ensuremath{j}\xspace}
\NewDocumentCommand{\pwd}{}{\ensuremath{p}\xspace} % as far as I can tell, this is never used
\NewDocumentCommand{\ucrit}{}{\ensuremath{U_{critical}}\xspace}
\NewDocumentCommand{\pval}{}{\ensuremath{p}\xspace}
\NewDocumentCommand{\origPwd}{}{\ensuremath{p_{\mathsf{old}}}\xspace}
\NewDocumentCommand{\newPwd}{}{\ensuremath{p_{\mathsf{new}}}\xspace}
\NewDocumentCommand{\finalPwd}{}{\ensuremath{p_{\mathsf{fnl}}}\xspace}
\NewDocumentCommand{\usedSegIdxSet}{}{\ensuremath{I}\xspace}
\NewDocumentCommand{\usedSegSet}{}{\ensuremath{S}\xspace}
\NewDocumentCommand{\origSeg}{}{\ensuremath{seg_{\mathsf{old}}}\xspace}
\NewDocumentCommand{\chainOne}{}{\ensuremath{chain_{\mathsf{1}}}\xspace}
\NewDocumentCommand{\chainTwo}{}{\ensuremath{chain_{\mathsf{2}}}\xspace}
\NewDocumentCommand{\newSeg}{}{\ensuremath{seg_{\mathsf{new}}}\xspace}
\NewDocumentCommand{\origExp}{}{\ensuremath{exp}\xspace}

\NewDocumentCommand{\origSegVec}{o}{\ensuremath{\vec{u}\IfNoValueF{#1}{_{#1}}}\xspace}
\NewDocumentCommand{\origExpVec}{o}{\ensuremath{\vec{z}\IfNoValueF{#1}{_{#1}}}\xspace}
\NewDocumentCommand{\newSegVecIntermediate}{o}{\ensuremath{\vec{w}\IfNoValueF{#1}{_{#1}}}\xspace}
\NewDocumentCommand{\newSegVec}{o}{\ensuremath{\vec{v}\IfNoValueF{#1}{_{#1}}}\xspace}
\NewDocumentCommand{\genNewSeg}{o}{\ensuremath{\mathit{newSeg}\IfNoValueF{#1}{\mathopen{(}{#1}\mathclose{)}}}\xspace}
\NewDocumentCommand{\genNewSegExp}{o o}{\ensuremath{\mathit{newSeg}\IfNoValueF{#1}{\mathopen{(}{#1}\IfNoValueF{#2}{,\,{#2}}\mathclose{)}}}\xspace}
\NewDocumentCommand{\boolAlter}{}{\ensuremath{\mathsf{alter}}\xspace}
\NewDocumentCommand{\genRelated}{o o}{\ensuremath{\mathit{generateRelated}\IfNoValueF{#1}{\mathopen{(}{#1}\IfNoValueF{#2}{,\,{#2}}\mathclose{)}}}\xspace}
\NewDocumentCommand{\concat}{o o}{\ensuremath{\mathit{concat}\IfNoValueF{#1}{\mathopen{(}{#1}, {#2}\mathclose{)}}}\xspace}
\NewDocumentCommand{\adversary}{o}{\ensuremath{A\IfNoValueF{#1}{\mathopen{(}{#1}\mathclose{)}}}\xspace}

\begin{document}
\title{\SysName: 
Using LLMs to Encourage 
Memorable and Secure Password Replacement}


\author{\IEEEauthorblockN{Anna Gerchanovsky}
	\IEEEauthorblockA{Duke University\\
		anna.gerchanovsky@duke.edu}
	\and
	\IEEEauthorblockN{Lujo Bauer}
	\IEEEauthorblockA{Carnegie Mellon University\\
		lbauer@cmu.edu}
	\and
	\IEEEauthorblockN{Michael K. Reiter}
	\IEEEauthorblockA{Duke University\\
		michael.reiter@duke.edu}}
% \author{Anonymous Authors}
	

%Telephone: (800) 555--1212, Fax: (888) 555--1212}


\IEEEoverridecommandlockouts
\makeatletter\def\@IEEEpubidpullup{6.5\baselineskip}\makeatother
\IEEEpubid{\parbox{\columnwidth}{
		Preprint
}
\hspace{\columnsep}\makebox[\columnwidth]{}}

\maketitle

\begin{abstract}
  We report on the
  design and evaluation of \sysName, a tool that helps a user replace
  her password when she is required to do so.  Left to their own devices, 
  users tend to replace their previous passwords with 
  predictable variations of the original ones. \SysName
  leverages LLMs to suggest password variations that are chosen to be
  easy for the user to remember but harder for an attacker to predict.
  To do this, \sysName infers the meaning behind original password
  components and then suggests semantically related 
  (yet syntactically unrelated)
  components for the new password.  In a user study, passwords created
  using \sysName were more secure than both replacement
  passwords created without using \sysName and original passwords. In
  particular, \sysName replacement passwords were harder to guess 
  in an online attack than
  alternative replacement passwords even by an attacker with knowledge
  of the original password and full knowledge of the tool
  implementation.
  Passwords created with \SysName were also comparably
  memorable to alternative replacement passwords and original passwords, 
  as measured by the ability of users to successfully log in a week
  after creating their password.
\end{abstract}

\section{Introduction}
\label{sec:intro}

In the context of managing one's security and privacy, password
maintenance is among the most famously difficult cognitive tasks for
humans to do well.  Even when informed about the leak of their
password information, a minority of users change their
passwords~\cite{Bhagavatula:2020:Breach}; if they do, a majority of
replacement passwords are weaker than the
originals~\cite{Bhagavatula:2020:Breach}; and few users change their
practices to better protect their accounts (e.g., adopt a password
manager~\cite{Ablon:2016:Attitudes} or enable two-factor
authentication~\cite{Thomas:2017:Risks}).  Only providing users with
understanding of \textit{both} the threat \textit{and} how to change
passwords was shown to increase the likelihood that users would do
so~\cite{Zou:2024:Encourage}.  It is thus not surprising that when
forced to change passwords proactively, to counter the hypothetical
possibility that their password information was leaked, users invest
little effort to make any changes but the obvious ones
(e.g.,~\cite{Zhang:2010:Expiration}).
As such, users resist changing passwords and, when they do, they
choose replacements that are insufficiently different from the
originals and therefore are easy to guess with knowledge of the
original passwords.
Tools like password managers can both store passwords and help create 
strong passwords; however, they can't be used in 
all contexts~\cite{Sadik:2025:Practices} and 
aren't used universally~\cite{Lyastani:2018:Managed,Zibaei:2022:Nudge}, 
and even in the presence of password managers, 
users need to create and remember some passwords, like the password to a password 
manager itself, themselves.

Considering the many ways in which large language models (LLMs) can be
useful in performing tasks that humans find cognitively
challenging~\cite{Brachman:2025:LLMs}, it might seem natural that LLMs
could play a useful role in password management, as well.  While
there have been some initial forays into doing so (see
\cref{sec:related} for a discussion of related work), we are aware of
none that leverage an LLM's ability to mimic human lexical
organization.  Indeed, a recent study using word
association~\cite{Xiao:2025:Lexicon} showed that modern LLMs model
human mental lexicons quite well.  The key contribution of this paper
is to show how this ability can be leveraged in a tool, called
\sysName, to assist a user with the cognitively difficult task of
replacing her password.

\SysName helps a user replace her password with a new one that is both
memorable and hard for an attacker to guess, even with knowledge of
the old password.  \SysName works by replacing all elements of the old
password with semantically similar but difficult to guess or predict
replacements.  Starting from the user's existing password (or her
initial attempt for her password), \SysName segments the password and
provides a semantic description for each segment.  For example,
starting from the password ``m1ndR3ADER'', the tool might identify
segments ``mind'' and ``reader'', explaining these as ``The human
brain or mental state and awareness'' and ``A person who reads written
content, such as text'', respectively.  The tool permits the user to
adjust the segments and explanations; e.g., she might delete the
segment ``reader'' and change the segment ``mind'' to ``mindreader'',
with the explanation ``able to know my thoughts''.  Finally, \sysName
generates chains of associated segments, combining selected ones into
candidate passwords, from which the user can choose (and adjust).  The
key insight in this technique is that since the user was presumably
able to recall her previous password, and because the associations
leading to new segments are ones that are likely meaningful to humans,
the user can recall the new password almost as easily.

In this paper, we shed light on the key design considerations to
realize this approach and how we instantiated these decisions in a
working tool.  We also compare \sysName to the unguided replacement
of passwords via a user study, yielding the following findings:
\begin{itemize}
\item \textbf{Memorability}: \SysName helps users create replacement
  passwords that are comparably memorable to both original passwords
  and their unguided replacements.

\item \textbf{Black-box security}: \SysName helps users create
  replacement passwords that are more secure against conventional guessing
  attacks---ones without knowledge of the password being
  replaced---than replacement passwords created without \SysName
  and the original password that is replaced.
\item \textbf{White-box Security}: \SysName helps users create
  replacement passwords that are more secure against online attacks
  \textit{designed with knowledge of \sysName} than replacement
  passwords created without \SysName, given an attacker \textit{with
    knowledge of the original password being replaced}.
\end{itemize}

For example, in our user study, we found that passwords created
with \SysName were \textit{never} guessed in online attacks, 
and only 58\% of these passwords were guessed in offline attacks,
compared to over 80\% of both alternative replacements and original
passwords. For an attacker with knowledge of the original 
password and the design of \SysName, only 3.5\% of \SysName passwords
were guessed compared to 18\% of alternative replacements in an online attack.
When users returned to log in, we found no statistically
significant difference in the login success rate between
users with \SysName passwords and those with alternative replacements
or original passwords.
Out of our user study participants who used \SysName, over 70\%
said they would use a password created with it.

The rest of this paper is structured as follows.  We discuss related
work in \cref{sec:related}. 
We discuss the implementation of \SysName in \cref{sec:tool}
and the design of our user study in \cref{sec:user}.
We lay out our methods for the evaluation of \SysName in \cref{sec:methods}
and go over the results in \cref{sec:results}.
We discuss the implications of our work 
and limitations in \cref{sec:discussion}.
We conclude our work and discuss ethical considerations in
\cref{sec:conclusion} and \cref{sec:ethics}. 

\section{Related Work}
\label{sec:related}

Helping a user improve the security of her chosen password as she is
deciding upon it is a goal that \sysName shares with other systems.
Password blocklists and composition policies~\cite{Tan:2020:Practical}
interfere with the selection of weak passwords; strength
meters~\cite{Ur:2012:Measure, Zimmermann:2023:Hybrid,
  Khernamnuai:2017:Context, Khernamnuai:2023:Augmenting, Doneva:2026:Encouraging}
and advice/nudges~\cite{Yildirim:2019:Encouraging, Guo:2020:Nudge,Doneva:2026:Encouraging}
encourage the selection of stronger ones.  The Telepathwords
system~\cite{Komanduri:2014:Telepathwords} shows ways that users would
often complete a password based on what the user has typed so far, to
enable the user to avoid these easily predicted completions.  And,
most closely related to our work, several techniques detect password
structure and advise the user how to strengthen
it~\cite{Ur:2012:Measure, Woo:2018:GuidedPass}.  \SysName improves
upon this last category of system by detecting not only structure but
meaning, and then suggesting specific alternatives based on that
meaning---e.g., to replace ``apple'' with ``pumpkin'' if the user
chose the former to mean ``an autumn fruit,'' but with ``touchscreen''
if the user chose the former to mean ``a consumer electronics
company''---instead of generic advice (e.g., ``Add an uncommon
word''~\cite{Woo:2018:GuidedPass}).

To our knowledge, in the context of passwords, LLMs have primarily
been studied as a technology to better model the distribution of
human-selected passwords, in order to guess such passwords more
efficiently (e.g.,~\cite{Rando:2023:PassGPT, Zou:2025:Guessing}). 
Out of the box, LLMs are known to create predictable, low-entropy
passwords~\cite{Irregular:2026:Vibe}.
Our method avoids this by shuffling segments, using 
chains of association to find replacement, and adding randomness
programmatically, and we evaluate security both against
an attacker with knowledge of the tool and its probabilities and 
one without this knowledge.
Li, et al.~\cite{Li:2025:LLM} explored the use of LLMs to generate
measurably high-entropy
passphrases for users but reported poor (essentially, no) memorability
after only three days in user studies; presumably this negative result
resulted, in part, from the fact that the passphrases were not
informed by the preferences of their participants.  In contrast, the
method we explore here produced considerably better memorability in
our user study, after one week.

More distantly related work focuses on preventing password reuse by
the same user at different sites through the use of different password
policies~\cite{Seitz:2017:Reuse} or explicitly detecting that
reuse~\cite{Wang:2021:Reuse}.
Also distantly related are previous works that have proposed methods
to improve how users replace existing passwords by ensuring that the
new password is not too syntactically similar to the previous
password(s) \textit{for the same account}, while storing only a
representation of the previous password(s) that makes them difficult
to recover in the event of a password database breach. For example,
Berardi, et al.~\cite{Berardi:2021:Probabilistic} propose a method
that leverages Bloom filters of password \gramLength-grams to strike a
balance between breach-resilience and the ability to detect similarity
of a new password to previous ones.  \SysName is consistent with such
solutions but does not require their use.  In particular, \sysName
does not impose nonstandard methods for storing password information.

Alternative types of passwords have also been explored to generate
memorable and secure passwords, including graphical passwords
(e.g.,~\cite{Jermyn:1999:Graphical, Davis:2004:Graphical,
  Dunphy:2007:DrawASecret}), password shapes for
PINs~\cite{Weiss:2008:PassShapes}, and replacing password recall with
word or letter recognition~\cite{Wright:2012:DoYouSee}.  \SysName is
designed specifically for text-based passwords that the user is
expected to recall---overwhelmingly the norm today---though extension
of its ideas to graphical passwords or other designs is an interesting
direction for future work.
\section{\SysName Design and Implementation}
\label{sec:tool}
In this section, we discuss how \SysName was implemented. 
We begin by covering the use case of \SysName in \cref{sec:tool:use_case}.
We then discuss the method used by \SysName to replace 
passwords in \cref{sec:tool:approach}
and finish by covering the general details of how \SysName was implemented
in \cref{sec:tool:setup}.

\subsection{Use Case}
\label{sec:tool:use_case}
A user may use \SysName
to change her password when asked or required to do so, potentially
in the case of a breach or password expiration, or at will.
\SysName is meant to be used to replace passwords by users for whom
password memorability is important, i.e., users who, at least in some
circumstances, choose to enter their passwords manually rather than
copy-pasting or auto-filling their passwords.  
This could be the case for devices that do not allow
password managers~\cite{Sadik:2025:Practices}, password-manager master passwords,
or any of the other passwords users choose to enter 
manually~\cite{Lyastani:2018:Managed}.
\SysName helps such a user change her password to one that is syntactically 
very different by analyzing the previous password and suggesting replacements 
whose components are semantically related to the components of the 
original password.

\subsection{Replacement Method}
\label{sec:tool:approach}
\SysName helps users pick better replacement passwords 
by replacing all relevant parts of the old
password with alternatives that are chosen to be easy to remember but
difficult to guess.  This process consists of identifying segments of
the original password, finding replacement segments for each, and
combining new segments into a replacement password.  In this section,
we describe these steps and illustrate them through screenshots from
\SysName.  More views of the user interface are shown in
\cref{sec:appendix:tool:ui}.

\subsubsection{Password Segmentation}
\label{sec:tool:approach:segmentation}
\begin{figure}[h]
    \centering
    \includegraphics[width=\linewidth, frame]{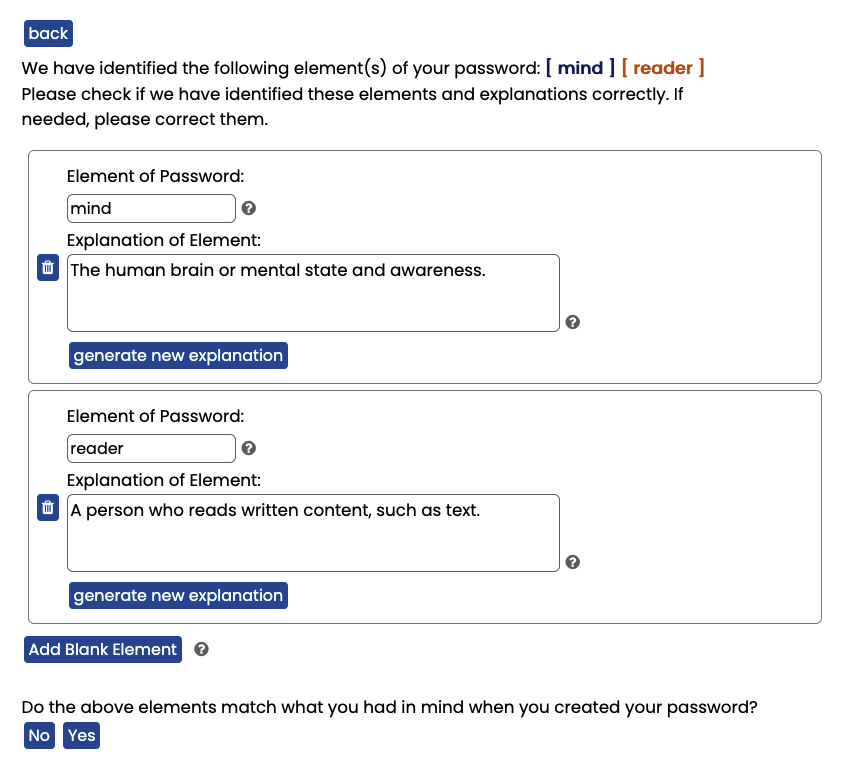}
    \caption{ A view of the segmentation
    stage of the tool for the original password ``m1ndR3ADER''.
    In the implementation of \SysName, we refer to ``segments'' as ``elements''.
    Users can edit elements or explanations and delete or
    add elements.
    }
    \label{fig:tool:segmentation}
\end{figure}

\SysName begins by identifying any segments of the password.  To
segment the password itself, \SysName uses
WordNinja~\cite{Anderson:2019:WordNinja}.  After piloting our study,
we found that it may be useful to attempt to undo any character
replacement, such as leetspeak, before segmentation.  \SysName uses an
LLM to identify whether or not passwords use leetspeak.  If so,
\SysName uses an LLM to attempt to undo this and recover the
underlying string.  With this method, for example, ``m1ndR3ADER''
becomes ``mindreader'', which can be segmented into ``mind'' and
``reader''.  While this method is not perfect, in our experience it
had few false positives (i.e., character replacements that were
unwarranted) and was able to recover passwords that most often
performed at least as well as the raw password when segmented.

After the password is segmented, \SysName generates a candidate explanation
for each, in order to identify what this segment may represent.
All prompts used by \SysName can be found in \cref{sec:appendix:tool:prompts}. 
Users are shown all segments and explanations.  They are given an
option to add new, blank segments, remove existing segments, or edit
existing segments by changing the segment itself or editing or
removing the explanation.  An example of the segmentation stage is
shown in \cref{fig:tool:segmentation} for the original password
``m1ndR3ADER''.  The user can edit any segments (``mind'' or
``reader'').  If she wants, for example, for those two segments to be
considered one, she could change ``mind'' to ``mindreader'' and delete
the ``reader'' segment using the trash icon. She could then either
enter her own explanation or click ``generate new explanation'' until
she is satisfied with the result. If she then wanted to add an
additional segment, she could click ``Add Blank Element'', which would
create a segment with empty fields she could then fill in.  Once she
is happy with the segmentation, she would click ``Yes'' at the bottom
of the screen. If she is unhappy and clicks ``No'', she will be
provided with additional information through tooltips, shown in
\cref{fig:appendix:tool:tooltips_segmentation}.

\subsubsection{Segment Replacement}
\label{sec:tool:approach:replacement}
\begin{figure*}[h]
    \centering
    \includegraphics[width=\linewidth, frame]{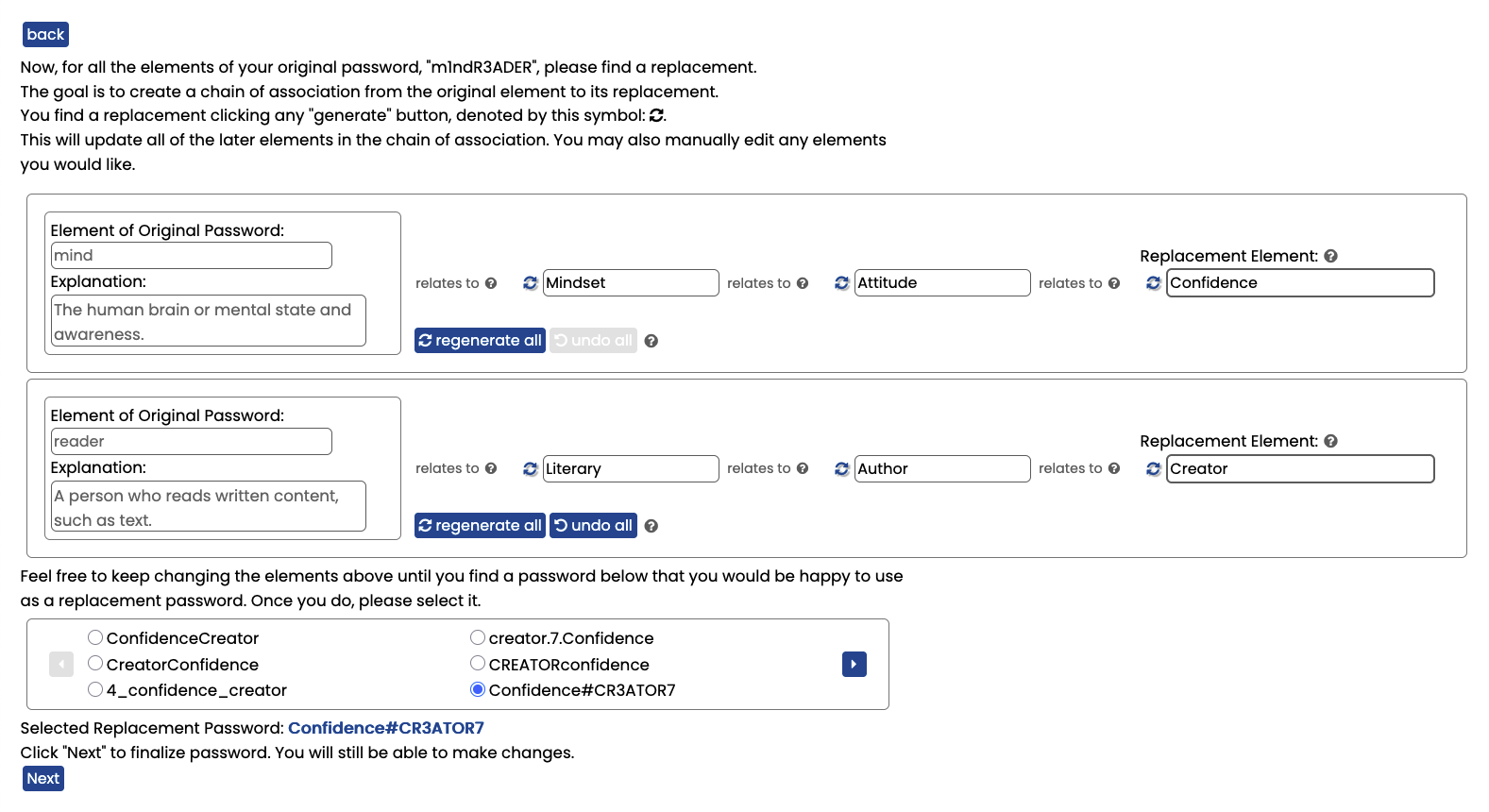}
    \caption{ A view of the segment-replacement
    stage of the tool. Here we again refer to segments as ``elements''.
    Users can edit the replacement element directly or by editing the chain,
    as well as select the replacement password.
    }
    \label{fig:tool:replacement}
\end{figure*}
\begin{algorithm}
  \caption{Algorithm for \genNewSeg}
  \label{alg:tool:approach:replacement_single}
  
  \KwIn{original segment \origSeg, a segment of \origPwd;
    an explanation \origExp for \origSeg.}
  \KwOut{\chainOne, \chainTwo, \newSeg}

  $\usedSegSet \gets \emptyset$ \\
  \tcc{elements of the chain that have been used}
  $\chainOne \gets \origSeg$ \\
  \While{$\chainOne \in \usedSegSet$} {
    $\chainOne \gets \genRelated[\origSeg][\origExp]$ \\
    \tcc{generates a related chain element using LLM}
  }
  $\usedSegSet \gets \usedSegSet \cup \{\chainOne\}$ \\
  $\chainTwo \gets \origSeg$ \\
  \While{$\chainTwo \in \usedSegSet$} {
    $\chainTwo \gets \genRelated[\chainOne]$ \\
  }
  $\usedSegSet \gets \usedSegSet \cup \{\chainTwo\}$ \\
  $\newSeg \gets \origSeg$ \\
  \While{$\newSeg \in \usedSegSet$} {
    $\newSeg \gets \genRelated[\chainTwo]$ \\
  }

\end{algorithm}

Once the segmentation is complete, each segment requires a
replacement. 
Not all segments are necessarily used in every replacement password
if the total length is sufficient, but we chose to set this requirement
to ensure more variance in the resulting replacement passwords. 
However, if a user does not want a replacement
for a given segment in their password, they can delete it in the previous
step of password segmentation.
A replacement is found by creating a chain of
association: From the original segment and explanation, an associated
string is generated.  From that string, another associated string is
found.  Finally, the replacement segment is generated from this
string.  In the end, \SysName shows the original segment, two
intermediate strings, and a replacement segment (the final element of
the chain). 
This process is described in \cref{alg:tool:approach:replacement_single}.
The chain of association method was used to add variance
to the replacement segment.  A chain length of three was selected in
order to make the resulting segment memorable, while adding variance.

\SysName also generates an explanation for the link between any two
elements of the chain, in order to make the process easier to
understand and interpret for users and increase memorability.  This
explanation is revealed in a pop-up to the user if the user hovers
over the ``relates to'' connector between the elements
(as shown in \cref{fig:appendix:tool:tooltip_link}).  

In \cref{fig:tool:replacement}, we show the segment replacement stage
where all segments have been replaced.  If a user would like to adjust the
replacement for some segment, she can do one of the following:
\begin{enumerate}
  \item Click ``regenerate all'', which would replace all items in the
    chain of association for that segment. For example, if she clicked
    this button in the top row, ``Mindset'', ``Attitude'', and
    ``Confidence'' would all be replaced.
  \item Click the regenerate button next to any element of the chain.
    This would cause that element and any later element to regenerate.
  \item Edit any element in the chain. Every later element would be
    regenerated. For example, if she changed ``Attitude'' to
    ``Thoughts'', the final element in the chain, ``Confidence'',
    would be regenerated.
  \item Edit the replacement element directly.
\end{enumerate}

\subsubsection{Password Creation}
\label{sec:tool:approach:creation}
\begin{figure}[h]
    \centering
    \includegraphics[width=\linewidth, frame]{
        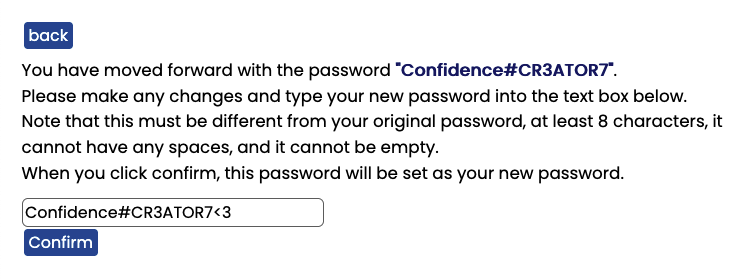}
    \caption{A view of the password-confirmation stage of the tool.
      In this case, the user added
      ``\raisebox{0.2ex}{$\scriptstyle<$}3'' to the selected password
      from \cref{fig:tool:replacement}.}  
    \label{fig:tool:selection}
\end{figure}
\begin{algorithm}
  \caption{Password Creation}
  \label{alg:tool:selection:method}
  
  \KwIn{vector \newSegVecIntermediate of replacement segments;
    Boolean \boolAlter}
  \KwOut{\newPwd}

  $\segsLen \gets 0$ \\
  $\usedSegIdxSet \gets \emptyset$ \\
  $\newSegVec \gets \emptyvec$ \\
  \For{$\segIdxIdx \gets 1$ \KwTo \vecLen{\newSegVecIntermediate}}{
    $\segIdx \getsr \nats[1][\vecLen{\newSegVecIntermediate}] \setminus \usedSegIdxSet$ \\
    $\usedSegIdxSet \gets \usedSegIdxSet \cup \{\segIdx\}$ \\
    $\newSegVec[\segIdxIdx] \gets \newSegVecIntermediate[\segIdx]$ \\
    \If{$(\segsLen \gets \segsLen + \strLen{\newSegVec[\segIdxIdx]}) > 18$}{
      \Break
    }
  }
  $\newPwd \gets \concat[\newSegVec][\boolAlter]$ \\
  \tcc{If \boolAlter, add leetspeak and capitalizations,
    and conjoin using `\_' or `.'}
\end{algorithm}
After replacing each segment of the original password, the replacement
segments are combined into candidate replacement passwords. 
The process by which the replacement segments are chosen and ordered is 
shown in \cref{alg:tool:selection:method}. The user
is shown six candidates at a time, and more are generated as the user
scrolls through sets of six.  This is shown at the bottom of
\cref{fig:tool:replacement}.

\SysName concatenates all or some of the replacement segments to
create \emph{permutation} passwords. 
To prevent the permutation passwords from being excessively long, 
we follow this process: If concatenating all segments results in a string 
less than 18 characters long, then this string is used as a permutation 
password. If concatenating all segments would result in a string of 18 or 
more characters, then we construct the permutation password by 
adding to it one segment at a time, and we stop the process as soon as 
we reach a length of 18 characters, potentially excluding some segments.

Additionally, candidate replacement passwords are generated
using segment permutation \emph{and} alterations.
For these passwords, all or some replacement segments are again 
used, following the same length criteria as before.
Each replacement segment is altered through randomized capitalization
styles (lowercase, uppercase, or first character uppercase), as well
as leetspeak-style character replacement.  Random numbers or special
characters are potentially added.  Replacement segments and additions
are combined either without separation through concatenation or with a
joining character (a period or an underscore).  Four out of the six
suggested passwords in \cref{fig:tool:replacement}, including the
selected password, are created with these alterations.
While alterations made passwords more difficult to guess,
we chose to allow users to select the simpler, permutation-only passwords
if they preferred, since they were more directly drawn from the replacement
segments.

The user selects a candidate replacement out of those shown.  After
selecting a replacement, the user is asked to type her final choice of
password.  She can either keep the chosen candidate unchanged or make
adjustments.  This step of the process is shown in
\cref{fig:tool:selection}.

Each step of the process is completed with user confirmation and with
an option for user input.  The user has an option to accept the
automatic completion of each step, to make the tool easier to use.
However, while all steps could be done fully automatically, requiring
user confirmation and allowing changes from the user allows the user
to better understand how the replacement password was created (with
the goal of increasing memorability), allows the user to select a
password that better matches her mental model (with the goal of
increasing memorability), and introduces additional variance to the
replacement password (with the goal of increasing security).

\subsection{Implementation}
\label{sec:tool:setup}
We hosted \SysName as a web application on our institution-controlled
server.  The web application used the Django framework.  We used two
large language models throughout the process, both Llama models. For 
time-sensitive responses we used Llama 3.2 with 3 billion parameters
(``Llama-3.2-3B-Instruct''), and for the rest we used Llama 3.1 with 8
billion parameters (``Llama-3.1-8B-Instruct'').  Both models were
instruction-tuned and downloaded from their official repositories on
HuggingFace.  They were run on the same institution-controlled server
that hosted the web application.

\section{User Study}
\label{sec:user}
To evaluate \SysName, we conducted a user study that was approved
by our Institutional Review Board.
We acknowledge that IRB approval is not necessarily sufficient 
to ensure principles of ethical research are followed, 
so we discuss how we set out to follow these principles in the user study in
\cref{sec:ethics:user} and our ethical considerations in general 
in \cref{sec:ethics}.
We recruited participants from Prolific with a gender-balanced sample.
The only requirement for this study was fluency in English.
Fully informed consent was obtained from all participants at the
start of the user study.
Before returning to later stages of the study,
users were provided access to the consent form for review.

\subsection{User Study Protocol}
\label{sec:user:protocol}
\begin{figure}[h]
    \centering
    \includegraphics[width=\linewidth]{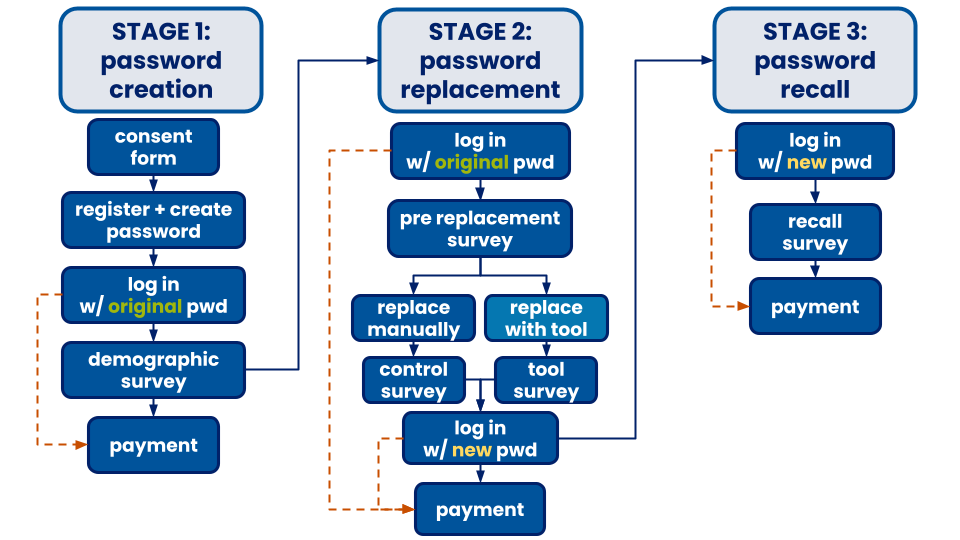}
    \caption{A flow chart describing the user study procedure.
    Blue arrows represent successful completion of a given step, while
    dashed orange arrows represent failure to login after 5 attempts,
    for which participants were also compensated.}
    \label{fig:user:protocol}
\end{figure}
We had two groups, which differed only for Stage 2.
We used Qualtrics for any surveys and the consent form.
All survey questions are included in \cref{sec:appendix:user:survey}
and participant demographics 
are included in \cref{sec:appendix:user:demographics}.

\subsubsection{Stage 1: Password Creation}
\label{sec:user:protocol:s1}
In Stage 1, users were recruited to the study. 
They filled out the consent form and were then directed to our website.
They were reminded of the process for the first stage.

Users registered for accounts using their 
Prolific IDs, so we could ensure payment,
 and passwords of 
their choice.
The password was required to be at least eight characters in length and have 
no spaces. Other than that, there were no restrictions.
Users had to enter their password on two separate fields during registration.
After registering, users were prompted to log in.

If users were unable to log in after five attempts, users were returned to 
Prolific and paid. 
After logging in, users were taken to a demographic survey.
After the survey, participants were returned to Prolific and compensated.
The compensation for this stage was \$1.

\subsubsection{Stage 2: Password Replacement}
\label{sec:user:protocol:s2}
A week after the first stage, participants from Stage 1 who were able to 
successfully complete the entire process were invited back for Stage 2.
We follow the example set in previous 
work~\cite{Taiabul:2017:Bits, Clark:2025:List}, where password 
recall is measured a week after password creation\footnote{Some 
additional papers measure at
1-5 days~\cite{Komanduri:2011:People,Shay:2012:Horse,Ur:2012:Measure,
Ur:2017:Driven,Tan:2020:Practical,Li:2025:LLM}.}.
As with Stage 1, users were reminded of what the process for the stage 
would be, and were provided with a link to a copy of the consent form.

In Stage 2, participants were asked to log back in with the password created
in the previous stage. If users failed to log in after five attempts, 
they were returned to Prolific and compensated.

If users were successful, they were taken to a survey. 
Participants were asked whether they wrote their password down 
the previous week, whether they typed their password or 
used copy-paste or autocomplete, and questions about how they came up with their 
password. After the survey, participants were told to replace their password
as if it had expired (see \cref{fig:appendix:tool:pre}).

One group of participants replaced their password without assistance,
which we refer to as \emph{manual replacement}.
We refer to these passwords as \emph{control passwords}.
The only additional restrictions on the new password 
were that of the old password
and the requirement that the new password be different.
After manually replacing their password, users were asked what method they used 
in a survey.

The other group used the tool, as described in \cref{sec:tool:approach}.
We refer to this group as \emph{tool replacement}.
These passwords had the same requirements as the manually replaced passwords.
After replacing their password with the tool, users were asked about
their experience using the tool.

After completing the corresponding post-replacement survey, 
both groups were asked to log back in with their new password.
If users failed after making five attempts, 
they were returned to Prolific and compensated.
If users succeeded, they were also returned to Prolific and compensated.

Users in the control group were compensated \$1.  Users in the tool
group were compensated \$3.  The difference in compensation was due to
the additional time required for tool replacement, since users had to
familiarize themselves with the tool.

\subsubsection{Stage 3: Password Recall}
\label{sec:user:protocol:s3}
A week after Stage 2, users who were able to complete the previous
stage completely---replacing their passwords and logging back in at
the end---were invited back for the final stage. Again, users were
told the process for this stage and provided access to a copy of the
consent form.

In this stage, users first needed to log in with the new password that
they created the previous week. If users made five attempts and failed,
they were returned to Prolific and compensated.

After logging back in, users were taken to a survey,
where they were asked how they remembered and entered their password. 
After this, users were returned to Prolific and compensated.
For this stage, users were compensated \$1.

\subsection{User Study Development}
\label{sec:user:development}
Before running our final study, we ran two pilot 
studies, both of which were fully IRB approved.
The first pilot consisted of 106 participants and followed
the same structure as the final study.
This pilot used a previous version of \SysName, 
which differed in the segment replacement step.
In that version, this step did not include a chain of association.
We found that passwords created with the previous version of \SysName
did not have the desired white-box security.

After implementing the chain of association to \SysName, we conducted
a brief mini-pilot with 52 participants.  In this mini-pilot, we aimed
to confirm the usability of \SysName before the final study.  This
study had only one stage, in which users created and replaced their
passwords.  This stage was a combination of Stage 1 and 2, including
only the demographic survey and post-replacement survey.  Participants
were compensated \$3.

We do not report on the memorability, security, or usability 
results from either pilot study in this work.

\subsection{User Study Summary}
\label{sec:user:summary}
We ran our final study in two batches, which began about a month apart.
For each batch, we recruited 100 participants.
Half of each batch was assigned to the control group and the
remaining half was in the tool group.
From our user study, we collected 201 old passwords
(due to an error, an additional participant was recruited),
61 control passwords, and 58 tool passwords.
\section{Evaluation Methods}
\label{sec:methods}
In this section, we describe how we evaluated \SysName.
First, we discuss the black-box and white-box security of our passwords in 
\cref{sec:methods:blackbox} and \cref{sec:methods:whitebox}.
Then, we discuss our measurements for 
password memorability in \cref{sec:methods:mem}.
Finally, we cover how we measured the usability of \SysName and tool passwords
in \cref{sec:methods:usability}.

\subsection{Black-Box Security}
\label{sec:methods:blackbox}
We separately analyzed the manually created (control) replacement passwords,
the tool replacement passwords,
and original passwords. For black-box security, we considered attacks
where the attacker does not have knowledge of the user's original password
or of \SysName's design.

Throughout the paper, we use guess estimates from 
Flor\^{e}ncio et al.~\cite{Florencio:2014:Admin}
to differentiate online and offline attacks.
We will refer to attacks up to $10^7$ guesses
as \textit{online} attacks ($10^7$ is an upper bound based on 
one guess per second for four months) and attacks up to $10^{20}$ guesses
as \textit{offline} attacks (based on 1,000 machines computing hashes
for a \emph{single} account for four months).
Other work uses thresholds such as $10^5$, $10^6$, $10^{10}$,
$10^{12}$, and $10^{14}$~\cite{Ur:2012:Measure,Oesch:2020:ThenNow,
Tan:2020:Practical,Lee:2022:Policies}.

\subsubsection{PGS}
\label{sec:methods:blackbox:pgs}
We used Carnegie Mellon University's Password Guessability Service (PGS),
which is available for researchers to use upon request,
to evaluate the black-box security of passwords~\cite{CMU:2025:PGS}.
PGS provided us with the expected number of guesses necessary to 
crack a password with the following approaches:
\begin{itemize}
\item \textbf{Hashcat:} Hashcat is a software tool that takes a word
  list (in this case, the ``CMU wordlist'') and transforms the words
  to crack passwords.

\item \textbf{John The Ripper (JTR):} John the Ripper is a popular
  open source password hash cracking tool. It is able to match common
  password hashes in order to crack the plaintext
  passwords~\cite{Openwall:2025:John, Marchetti:2022:JohnAnalysis}.
  It also uses a wordlist and word transformations.

\item \textbf{Markov:} PGS uses code from Ma, et
  al.~\cite{Ma:2014:Markov}.  The implementation used is a
  5\textsuperscript{th}-order Markov model.

\item \textbf{Probabilistic Context-Free Grammar (PCFG):} Weir, et al.
  first introduced using probabilistic context-free grammar for
  password cracking~\cite{Weir:2009:ContextFree}.  PCFG generates
  password guesses using context-free grammar and probabilities from a
  training set of existing passwords.  This method has been shown to
  be more effective than John the
  Ripper~\cite{Kelley:2012:GuessAgain}.  The implementation used by
  PGS is developed by Komanduri~\cite{Komanduri:2018:PCFG}.

\item \textbf{Neural Network:} The neural network approach is
  described by Melicher, et al.~\cite{Melicher:2016:Fast}.  Neural
  networks can outperform Markov and PCFG, especially at high guess
  counts.
\end{itemize}
All methods but the last were evaluated by Ur, et
al.~\cite{Ur:2015:RealWorld}.

We selected the `Length 8+' password policy in PGS, as that was a
requirement in our user study.  We used the recommended option for all
approaches.  In our results, we report on the minimum number of
guesses necessary to crack a password across the five approaches.

\subsubsection{PassGPT}
\label{sec:methods:blackbox:passgpt}
Alternative methods to crack a password have included training
generative models, like a GAN (generative adversarial network) or LLM,
where success is measured by the fraction of passwords generated in
some number of guesses from a unique password test
set~\cite{Pasquini:2021:Representation, Rando:2023:PassGPT}.  We used
PassGPT~\cite{Rando:2023:PassGPT}, an LLM-based password generator, to
see how many passwords were cracked, and the generation loss with
PassGPT.  We generated $10^8$ passwords with PassGPT and calculated
which fraction of original passwords, control passwords, and final
tool passwords were guessed in some amount of guesses.  Additionally,
we calculated the total PassGPT loss over all passwords to estimate
the likelihood they would be guessed, as many passwords were not
guessed in $10^8$ guesses.

\subsection{White-Box Security}
\label{sec:methods:whitebox}
For white-box security, we considered an attacker who has knowledge of
the old password, as well as full knowledge of \SysName, 
where applicable, and aims to learn the replacement password. 
We used the same guess cutoffs for online versus offline attacks 
as we did for black-box security.

First, we considered password overlap as in previous papers, 
as measured by \textit{consecutive} character overlap. This is
a proxy for how difficult a password would be to guess.  In previous
work, passwords overlapping in less than three consecutive
characters were considered ``completely
dissimilar''~\cite{Bhagavatula:2020:Breach}.

Attacks have been developed to leverage a user's password or passwords
at other sites, measuring the similarity of 
a new password based on edit distance 
from a known password
and edit probability and generating guesses 
with encoder-decoder models 
\cite{Xiu:2024:PointerGuess, Wang:2023:Pass2Edit, 
Pal:2019:Pass2Path, He:2022:Passtrans}.
We used one of these methods, \PassToPath \cite{Pal:2019:Pass2Path},
to approximate the probability of guessing a replacement 
password given the original password.

Next, we aimed to more concretely 
evaluate how difficult it would be to crack the password.
We used the best of the methods in \cref{sec:methods:blackbox:pgs}, \PassToPath,
and our own ``knowledgeable attacker'' strategy.
This final strategy assumes the attacker has knowledge of (1) the password,
(2) the user's segmentation 
(but not the user's explanations for the original segments),
and (3) full knowledge of the tool implementation.

We combined all methods by choosing the minimum number of guesses required
between methods (PGS, \PassToPath, and the knowledgeable
attacker method described in \cref{sec:methods:whitebox:knowledgeable}
and \cref{sec:methods:whitebox:selected}).
Since \PassToPath and the knowledgeable attacker strategy result 
in probabilities of guessing correctly,
we took their inverse to get 
the expected number of guesses necessary.

\subsubsection{Knowledgeable Attacker Strategy}
\label{sec:methods:whitebox:knowledgeable}
\begin{algorithm}
  \caption{Adversary's strategy}
  \label{alg:methods:whitebox:method}
  
  \KwIn{vector \origSegVec of segments for original password \origPwd;
    Boolean \boolAlter}
  \KwOut{\newPwd}

  $\segsLen \gets 0$ \\
  $\usedSegIdxSet \gets \emptyset$ \\
  $\newSegVec \gets \emptyvec$ \\
  \For{$\segIdxIdx \gets 1$ \KwTo \vecLen{\origSegVec}}{
    $\segIdx \getsr \nats[1][\vecLen{\origSegVec}] \setminus \usedSegIdxSet$ \\
    $\usedSegIdxSet \gets \usedSegIdxSet \cup \{\segIdx\}$ \\
    $\newSegVec[\segIdxIdx] \gets \genNewSeg[\origSegVec[\segIdx]]$ \\
    \tcc{uses LLM to generate replacement segment}
    \If{$(\segsLen \gets \segsLen + \strLen{\newSegVec[\segIdxIdx]}) > 18$}{
      \Break
    }
  }
  $\newPwd \gets \concat[\newSegVec][\boolAlter]$ \\
  \tcc{If \boolAlter, add leetspeak and capitalizations,
    and conjoin using `\_' or `.'}
\end{algorithm}

The adversary uses the method in \cref{alg:methods:whitebox:method} to
try to guess the replacement password given the old password
segmentation and knowledge of the tool.

We call the adversary's guess $\adversary[\origSegVec]$.
We
approximate the probability that the adversary guesses the
password \newPwd (i.e., $\adversary[\origSegVec]=\newPwd$) using
\begin{align*}
\Pr(\adversary[\origSegVec] = \newPwd \mid \boolAlter)
    &= \sum_{\newSegVec} \Pr(\newSegVec \mid \origSegVec)
       \Pr(\newPwd \mid \newSegVec \wedge \boolAlter) \\
\Pr(\adversary[\origSegVec] = \newPwd \mid \neg \boolAlter)
    &= \sum_{\newSegVec} \Pr(\newSegVec \mid \origSegVec)
       \Pr(\newPwd \mid \newSegVec \wedge \neg\boolAlter)
\end{align*}
where here, \newSegVec denotes the multiset as passed to \concat.

We estimate $\Pr(\newSegVec \mid \origSegVec)$ by invoking
\genNewSeg[\origSegVec[\segIdx]] for each $\segIdx \in
\nats[1][\vecLen{\origSegVec}]$ repeatedly.  Specifically, we invoke
\genNewSeg[\origSegVec[\segIdx]] $10{,}000$ times
and set $\Pr(\newSegVec[\segIdx] \mid \origSegVec[\segIdx])$ to be the
fraction of these attempts that yielded \newSegVec[\segIdx].  For any
value that is never output, we set $\Pr(\newSegVec[\segIdx] \mid
\origSegVec[\segIdx])$ to $1/10{,}001$.  This means that we conservatively
assume that each new segment is guessed from the corresponding
old segment in the next try.  
Finally, we set $\Pr(\newSegVec \mid \origSegVec) \gets
\prod_{\segIdx\in\nats[1][\vecLen{\origSegVec}]}
\Pr(\newSegVec[\segIdx] \mid \origSegVec[\segIdx])$.

We calculate $\Pr(\newPwd \mid \newSegVec \wedge \boolAlter)$ and
$\Pr(\newPwd \mid \newSegVec \wedge \neg\boolAlter)$ explicitly.
Finally, we take $\Pr(\boolAlter) = \Pr(\neg\boolAlter) = 1/2$, and
put these values together to estimate $\Pr(\adversary[\origSegVec]) =
\newPwd$.

\subsubsection{Final Replacement Password vs.\ Selected Replacement Password}
\label{sec:methods:whitebox:selected}
In some cases, the final password used in the tool group is impossible
to get using \SysName's method of combining new segments.  This could happen,
for example, if the user adds more segments or if the user replaces
characters in new segments in a way that we do not. The attacker can
still guess the replacement password that was selected by the user,
which must be possible to get with \SysName.  We call this password
\finalPwd, and we now write the probability that the attacker guesses
a $\finalPwd \not\in \Supp{\adversary[\origSegVec]}$, where
$\Supp(\adversary[\origSegVec])$ denotes the support of the
distribution \adversary[\origSegVec], as
\begin{align*}
\lefteqn{\Pr(\adversary[\origSegVec] = \finalPwd)} \\
& = \max_{\pwd\in \Supp(\adversary[\origSegVec])} \Pr(\adversary[\origSegVec]=\pwd \wedge \PassToPath(\pwd)=\finalPwd) \\
& = \max_{\pwd\in \Supp(\adversary[\origSegVec])} \Pr(\adversary[\origSegVec]=\pwd) \Pr(\PassToPath(\pwd)=\finalPwd) \\
& \approx \Pr(\adversary[\origSegVec]=\newPwd) \Pr(\PassToPath(\newPwd)=\finalPwd)
\end{align*}
where \newPwd denotes the password selected by the user before
changing it into \finalPwd.

\subsection{Memorability}
\label{sec:methods:mem}
We measured the time to log in,
number of attempts,
and ability to log in.
There is one registration (during Stage 1),
two logins with the original password (during Stages 1 and 2),
and two logins with the replacement password (during Stages 2 and 3).
We compared: the memorability of tool passwords during Stage 2
to control passwords during Stage 2 and
original passwords during Stage 1 (immediately after passwords are created);
and the memorability of tool passwords during Stage 3
to control passwords during Stage 3 and
original passwords during Stage 2 
(one week after the passwords were created).

We separately report results for our entire set of participants,
our set of participants excluding those who either used auto-complete
or pasted their passwords based on keystrokes,
and our set of participants excluding those who used auto-complete,
pasted their passwords, or self-reported writing down
or pasting their passwords in our study.

\subsection{Tool Usability}
\label{sec:methods:usability}
We evaluated the usability of \SysName itself 
in \cref{sec:methods:usability:satisfaction}. 
We evaluated the usability of passwords created with \SysName
by measuring the rates at which passwords were manually entered 
in \cref{sec:methods:usability:pasting}
and login time in \cref{sec:methods:usability:time}.

\subsubsection{Tool Satisfaction}
\label{sec:methods:usability:satisfaction}
We evaluated participants' satisfaction with \SysName with a survey,
completed by participants in the tool group in Stage 2 of the study.
Participants responded to the following questions with a 5-point Likert scale:
\begin{itemize}
    \item I would use a password created by this tool.
    \item The tool was difficult to use.
    \item The tool was fun to use.
    \item The tool was intuitive to use.
\end{itemize}
We refer to these questions as `would use', `difficult', `fun', `intuitive'
in \cref{sec:results:usability:satisfaction}, respectively.

\subsubsection{Pasting Rates}
\label{sec:methods:usability:pasting}
For all login events mentioned in \cref{sec:methods:mem},
we checked whether the user entered her password with auto-complete
or by copy-pasting.
We say the user used auto-complete if no keystrokes were recorded,
and the user pasted her password if only one keystroke was recorded.
In our analysis, we only considered the two categories:
manually typed and pasted, which includes both copy-pasting and autocomplete.

\subsubsection{Login Times}
\label{sec:methods:usability:time}
For all login events, we evaluated time it took to log in, excluding
cases where users were unable to log in
or pasted their passwords.  We measured three types of time:
\begin{enumerate}
    \item \textbf{Overall Time:} We measure time from the first password 
    keystroke to the time the login was successful.
    \item \textbf{Typing Time:} We measure time from the first password 
    keystroke to the last keystroke. 
    If the user made multiple attempts to log in, 
    typing time is the time from the final, successful attempt.
    \item \textbf{Time-per-character:} We report
    typing time divided by the number of characters in the password.
\end{enumerate}

\section{Results}
\label{sec:results}
This section mirrors the structure of \cref{sec:methods}.
We begin by covering our results for password security:
first black-box security in \cref{sec:results:blackbox} and 
second white-box security in \cref{sec:results:whitebox}.
We then cover password memorability in \cref{sec:results:mem}
and finish by covering usability in \cref{sec:results:usability}.

\subsection{Black-Box Security}
\label{sec:results:blackbox}
\begin{figure}[h]
    \centering
    \includegraphics[width=\linewidth]{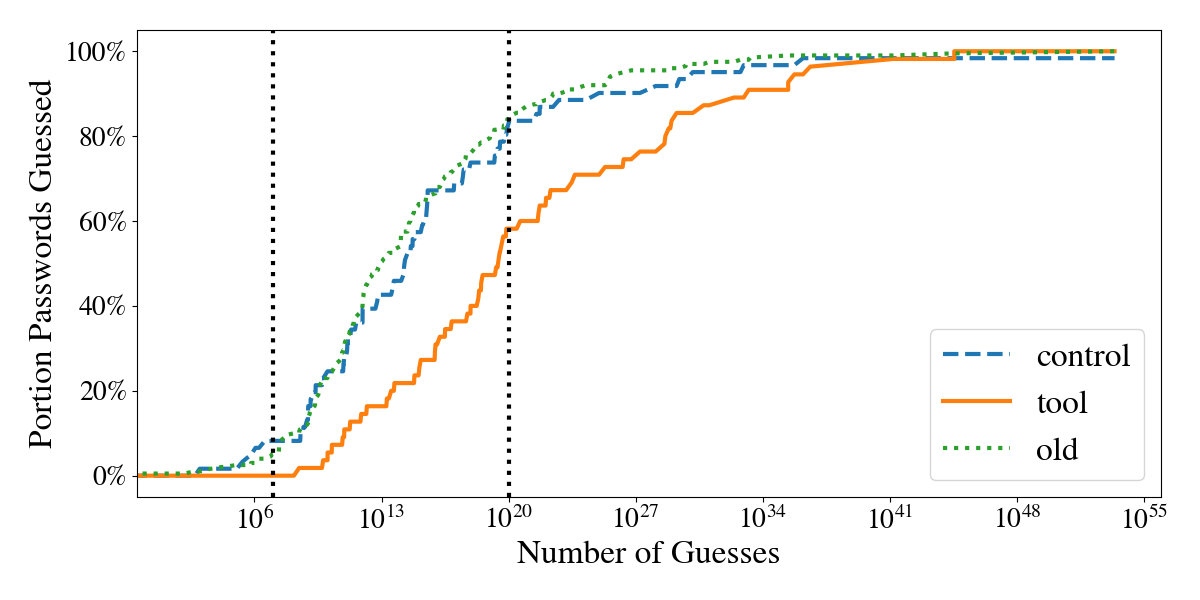}
    \caption{Portion of passwords
    guessed within a number of guesses
    for black-box PGS password cracking strategies, if using the minimum
    number of guesses between all approaches. 
    There are vertical dotted lines at $10^7$ and $10^{20}$ guesses,
    denoting cutoffs for online and offline attacks.
    Until about $10^{43}$ guesses,
    fewer tool passwords were always guessed than either control or
    old passwords, indicating that they are more secure against black-box
    attacks.
    }
    \Description{Tool Passwords harder to guess}
    \label{fig:results:pgs}
\end{figure}
For every method of measuring black-box security we evaluated, tool passwords
performed better than control passwords.
\subsubsection{PGS}
\label{sec:results:pgs}
\begin{table}[!ht]
    \centering
    \begin{tabular}{@{}r@{\hspace{1.5em}}r@{\hspace{0.5em}}r@{\hspace{0.5em}}r@{\hspace{1.5em}}r@{\hspace{0.5em}}r@{\hspace{0.5em}}c@{}}
      \toprule
      & \multicolumn{3}{c}{\textbf{online}} & \multicolumn{3}{c}{\textbf{offline}} \\
      & old & control & tool & old & control & tool \\
      \midrule
        Hashcat & 0.04\% & 0.03\% & 0.00\% & 21.50\% & 19.67\% & ~3.64\%\\
        JTR & 0.04\% & 0.03\% & 0.00\% & 31.50\% & 32.79\% & ~7.27\%\\
        Markov & 0.04\% & 0.03\% & 0.00\% & 13.50\% & ~9.84\% & ~5.45\%\\
        PCFG & 0.04\% & 0.03\% & 0.00\% & 47.00\% & 39.34\% & 12.73\%\\
        Neural & 0.04\% & 0.07\% & 0.00\% & 84.00\% & 81.97\% & 58.18\%\\
        Minimum & 0.05\% & 0.08\% & 0.00\% & 84.00\% & 81.97\% & 58.18\%\\
        \bottomrule
    \end{tabular}
    \caption{
    Portion of passwords guessed by various black-box PGS password cracking 
    strategies. 
    For every comparison, fewer tool passwords are guessed than
    control or old passwords.
    A password is guessed online if it takes under $10^7$ 
    guesses, and it is guessed offline if it takes under $10^{20}$ guesses.
    }
    \label{tab:results:guesses}
\end{table}
Between PGS methods, tool passwords
were always guessed later than control or original passwords.
Original passwords and control passwords were generally comparable.
The minimum number of guesses necessary 
to crack a password across all approaches 
is shown in \cref{fig:results:pgs},
and portion of passwords guessed in online
and offline attacks separated by approach
are shown in \cref{tab:results:guesses}.

Four passwords were not evaluated by PGS, since they contained non-ascii 
characters. One of them was an old password, and the rest were passwords
created with the tool. These passwords are omitted from this analysis.
Additionally, one password, from the control group replacement password set,
was never guessed across approaches. We set the number of guesses for this 
password to the largest number of guesses seen across passwords, 
$2\times 10^{53}$.

We examined whether the necessary log (base 10) 
number of guesses differed across groups
(old, control, and tool passwords) with a one-way analysis of variance (ANOVA). 
We found a significant effect, $F = 15.82, \pval < 10^{-7}$.
We performed ad-hoc analysis with Tukey's HSD, and found that 
the tool passwords were statistically different from both the 
old (mean difference 5.4 and $\pval = 0.0004$) and 
control (mean difference 6.4 and $\pval < 10^{-7}$) passwords.
There was no statistical difference between the old and control group.

\begin{figure}[h]
    \centering
    \includegraphics[width=\linewidth]{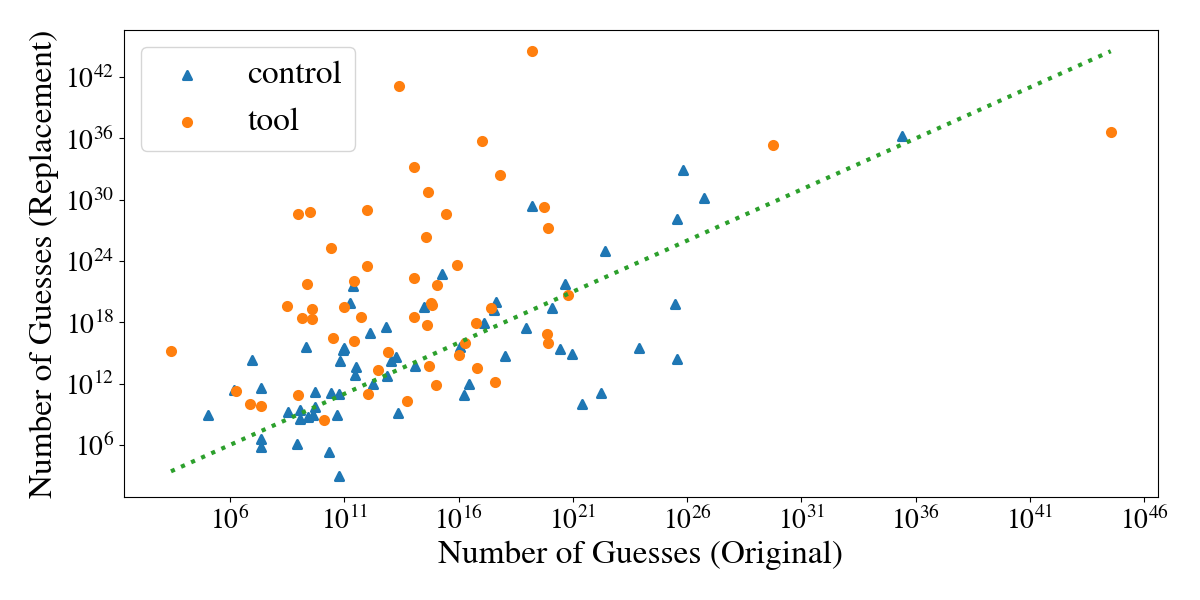}
    \caption{
    Comparison of the number of guesses required to guess
    old and replacement passwords, based on the smallest number
    of guesses among black-box PGS approaches.
    The green dotted line is $x=y$, so dots above the line represent
    cases where replacement passwords were harder to guess and points 
    below the line represent cases where replacement passwords were
    easier to guess.
    Tool passwords were more likely to show improvement than control passwords.
    }
    \label{fig:results:pairwise}
\end{figure}
We also compared the change in security, as evaluated by PGS,
between the old and replacement passwords.
\cref{fig:results:pairwise} compares the number of guesses required
to guess the old password and the corresponding replacement.
On average, control passwords took 2.2 times as long to guess 
as the original. Tool passwords, however, took $10^{6.8} \times$ 
as long. 43.33\% of control passwords were actually
weaker---easier to guess---than the original password. 
For tool passwords, only 23.64\% were easier to guess.

\subsubsection{PassGPT}
\begin{figure}[h]
    \centering
    \includegraphics[width=\linewidth]{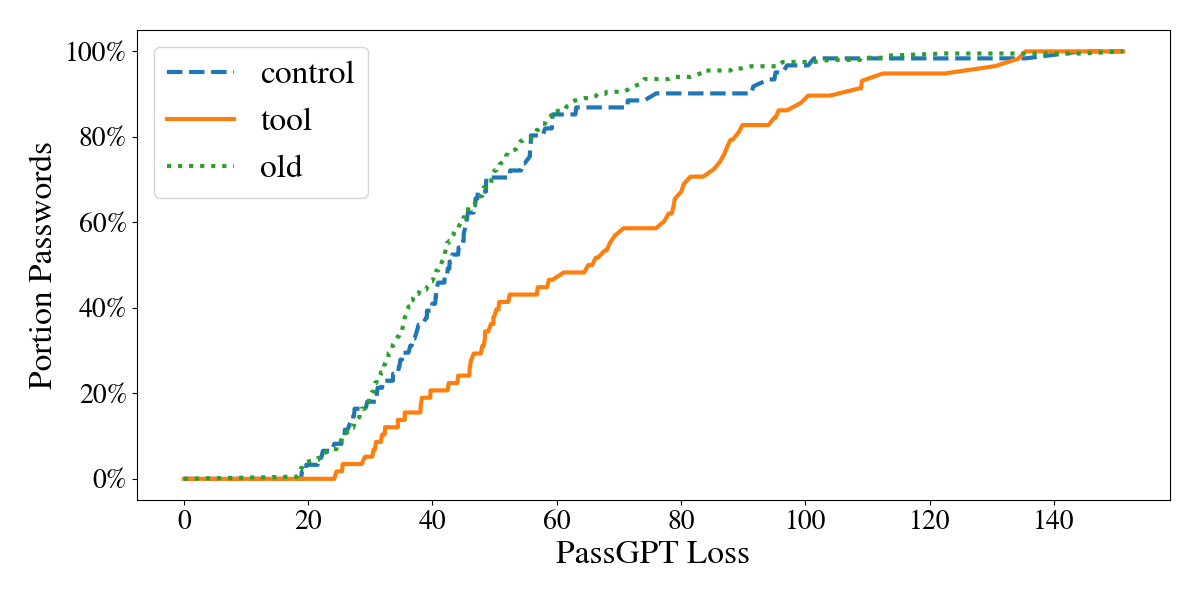}
    \caption{Portion of passwords with up to a given black-box PassGPT loss.
    Fewer tool passwords are guessed at any loss, up to about 120,
    meaning they tend to have higher loss than both control and old passwords.}
    \label{fig:results:passgpt:loss}
\end{figure}
Most passwords were not guessed in the $10^8$ PassGPT guesses we made.
Only 9/201 old passwords were guessed and 4/61 control passwords were.
No tool passwords were ever guessed.

Since so few passwords were guessed, we also measured PassGPT loss over
all passwords, as seen in \cref{fig:results:passgpt:loss}. 
Tool passwords consistently had higher loss
than original or control passwords, indicating they were
less likely to be generated by PassGPT.
Control passwords had slightly higher loss
than original passwords.

\subsection{White-Box Security}
\label{sec:results:whitebox}

\subsubsection{Password Overlap}
\label{sec:results:condsec:overlap}
\begin{figure}[h]
    \centering
    \includegraphics[width=\linewidth]{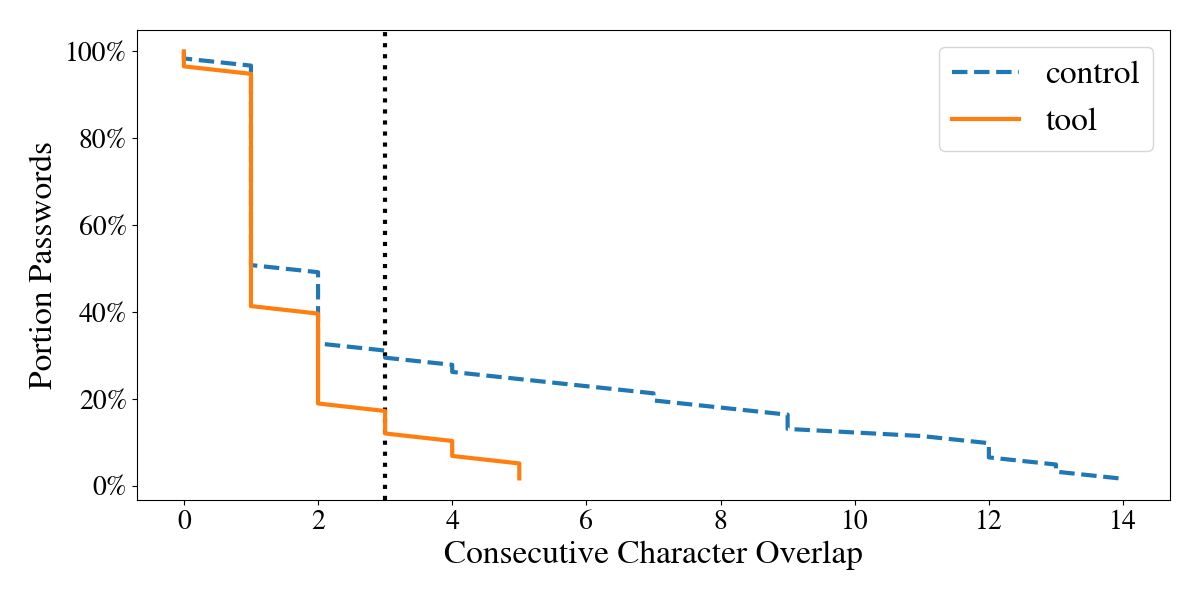}
    \caption{Consecutive character overlap between 
    original and replacement password. Figure shows portion of passwords
    with at least a certain number of character overlap.
    At any number of characters, either equal or fewer tool passwords
    than control passwords have given overlap with the corresponding original 
    password, indicating that tool passwords are more resistant to white-box
    attacks where the attacker knows the original password.
    The vertical dotted line represents three character overlap,
    less than which passwords are considered dissimilar.
    }
    \Description{Tool Passwords overlap less}
    \label{fig:results:overlap_char}
\end{figure}
\begin{figure}[h]
    \centering
    \includegraphics[width=\linewidth]{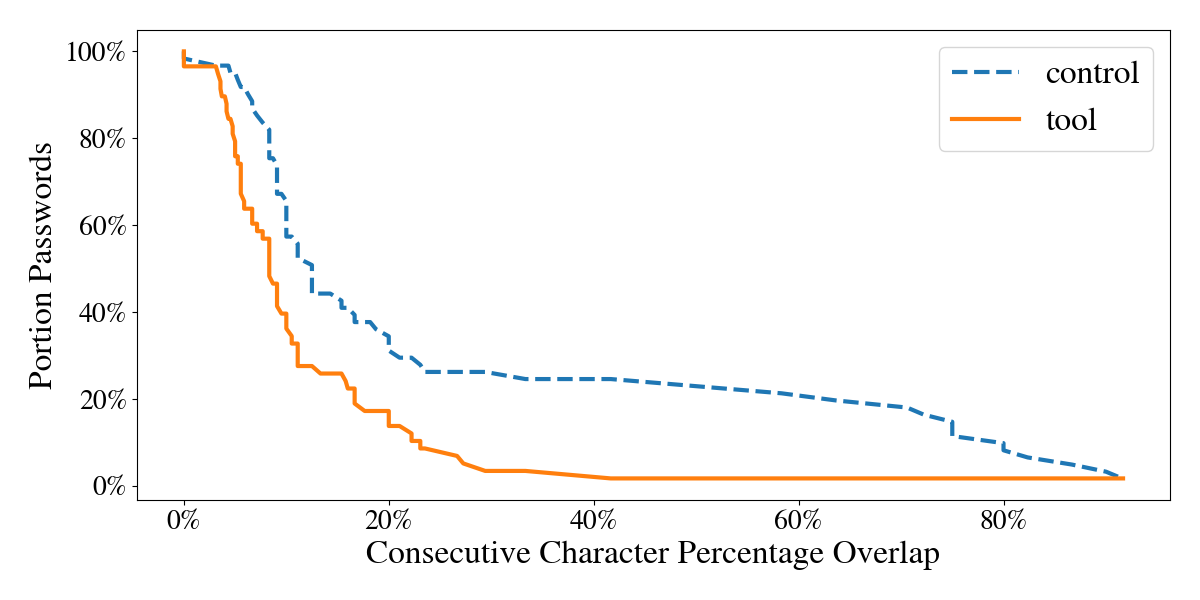}
    \caption{Consecutive character percentage overlap between 
    original and replacement password. 
    Percent is determined from the length of the longer password of the pair.
    For example, the passwords ``HelloWorld'' and ``Cello'' have 4 character 
    overlap (``ello''), which is 40\% of the length of
    the longer password of the two.
    Figure shows portion of passwords
    with at least a percentage overlap.
    Fewer tool passwords have at least a given overlap, for any overlap,
    meaning tool passwords overlap less with their original password than
    control passwords do, indicating that tool 
    passwords are more resistant to white-box
    attacks where the attacker knows the original password.
    }
    \Description{Tool Passwords overlap less}
    \label{fig:results:overlap_percent}
\end{figure}
There was less consecutive character overlap in the tool passwords than the
control passwords. 
If passwords with less than 3 character overlap are considered completely 
dissimilar, 69\% of control passwords were completely dissimilar
from the original, while 83\% of tool passwords were completely dissimilar.
All character overlap is shown in \cref{fig:results:overlap_char}.
Control passwords had, on average, 26\% overlap 
with the original password.
Tool passwords had average overlaps of 10\%.
Overlap percentages are shown in \cref{fig:results:overlap_percent}.

\subsubsection{Best White-Box Guessing Probability}
\begin{table}[!ht]
    \centering
    \begin{tabular}{@{}r@{\hspace{1em}}r@{\hspace{1em}}r@{\hspace{1em}}r@{\hspace{1em}}r@{\hspace{1em}}c@{}}

        \toprule
            &\multicolumn{2}{c}{\textbf{portion guessed}} 
            &\multicolumn{2}{c}{\textbf{average log guesses}} \\
            & \multicolumn{1}{c}{online} 
            & \multicolumn{1}{c}{offline} 
            & \multicolumn{1}{c}{online} 
            & \multicolumn{1}{c}{offline} \\
        \midrule
            control & 18.03\%$^\dagger$ & 88.52\% & 4.79$^\ddagger$ & 11.60 \\ 
            tool & 3.45\%$^\dagger$ & 81.03\% & 6.69$^\ddagger$ & 11.68\\ 
        \bottomrule
    \end{tabular}
    \caption{Comparison of tool and control passwords against online 
    and offline white-box attacks. 
    A password is guessed online if it takes under $10^7$ 
    guesses, and it is guessed offline if it takes under $10^{20}$ guesses.
    $\ddagger$ denotes $\pval < 0.005$ and $\dagger$ denotes $\pval < 0.01$.
    For values not labeled with $\dagger$ or $\ddagger$, 
    results are not statistically significant ($\pval > 0.1$).
    Tool passwords perform statistically significantly better, meaning they
    are harder to guess, than control passwords for online attacks, 
    while the difference is statistically insignificant for offline attacks.
    }
    \label{tab:results:whitebox}
\end{table}
\label{sec:results:condsec:overall}
\begin{figure}[h]
    \centering
    \includegraphics[width=\linewidth]{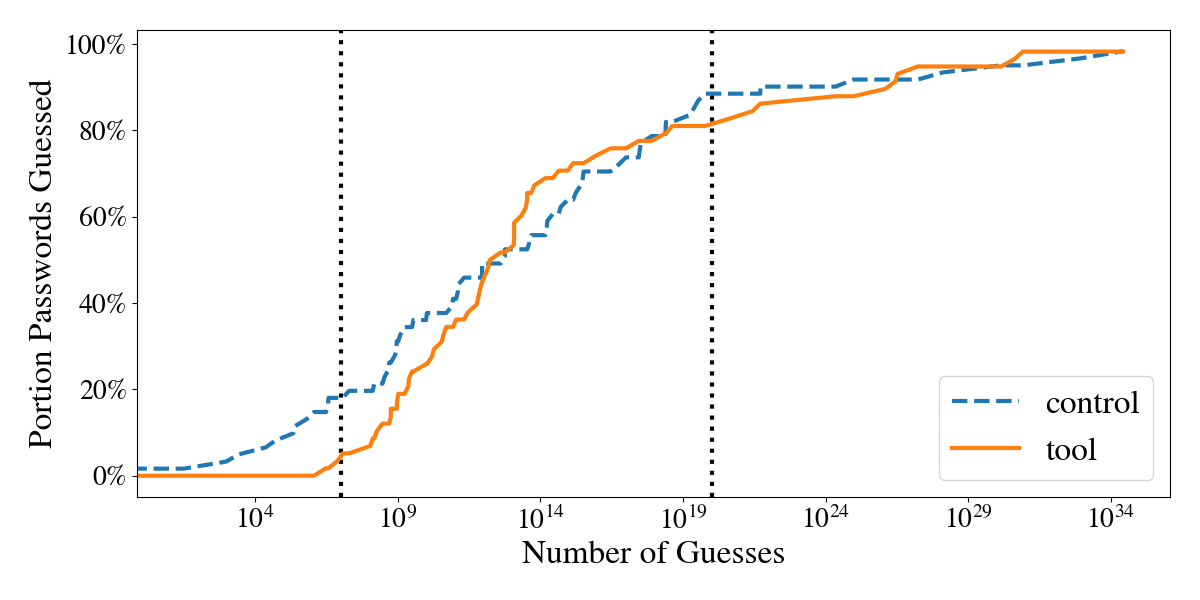}
    \caption{Portion of replacement passwords that were
    guessed in a white-box attack within a given number,
    given knowledge of the original password and \SysName.
    There are again vertical dotted lines at $10^7$ and $10^{20}$ guesses,
    denoting cutoffs for online and offline attacks.
    The guess number for any password is the minimum between 
    the PGS guess number, the inverse of \PassToPath probability,
    and the inverse of the knowledgeable attacker probability for tool
    passwords.
    Fewer tool passwords are guessed at most guess numbers,
    other than between about $10^{13}$ to $10^{18}$ and after $10^{28}$.
    }
    \Description{Tool Passwords always have a lower probability}
    \label{fig:results:best_cond}
\end{figure}
\begin{figure}[h]
    \centering
    \includegraphics[width=\linewidth]{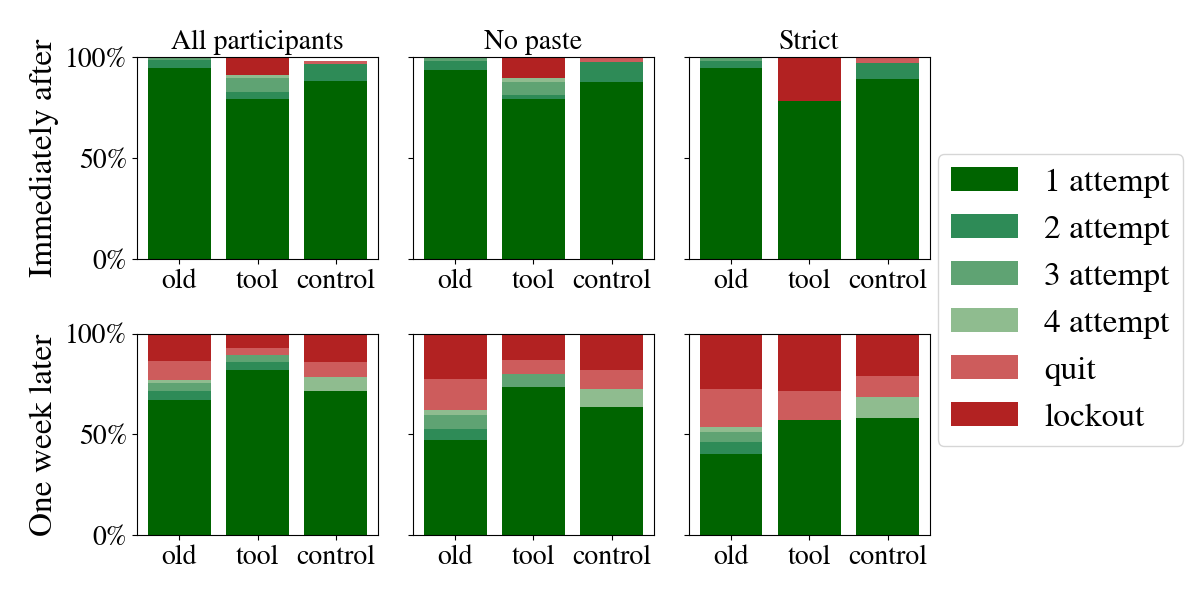}
    \caption{
        Login success rates,
        separated by password type, exclusion criteria, and time of login.
        No participants are excluded for ``All participants''.
        Participants who pasted or auto-completed their password,
        based on keystroke count, are excluded for ``No-paste''.
        For ``Strict'', all participants excluded in ``No-paste'' are excluded,
        as well as participants who reported writing down or copying
        their password in the survey. 
        Green colors represent success, with lighter
        colors representing more attempts required. Red colors represent
        either lockout (making 5 login attempts that all fail)
        or quitting (beginning to type in the password but not getting locked out).
        Immediately after password creation, tool passwords do worse.
        However, a week later, the difference between groups is statistically
        insignificant.
    }
    \label{fig:results:mem}
\end{figure}
Control passwords were easier to guess than the
tool passwords by a white-box attacker
(even though the white-box attacker had three strategies for 
tool passwords---PGS, \PassToPath, and the knowledgeable
strategy---but only two strategies for control passwords).
\cref{fig:results:best_cond} shows the probability of guessing 
tool and control passwords, 
given knowledge of the original password and \SysName.
49/58 of tool passwords 
were guessed first by an attacker cracking the password
with knowledge of the tool.
1/58 of these passwords were guessed first with Pass2Path.
The rest were guessed first using PGS,
without knowledge of the original password.
16/61 of control passwords were first guessed with Pass2Path.
The remainder were guessed first with PGS.
We did not attack control passwords 
with an attacker with knowledge of the tool. 

For online attacks, 3.45\% of tool passwords compared to 18.03\% of 
control passwords were guessed. 
We used a one-tailed Mann-Whitney U test which showed statistical significance
($\pval = 0.0056$, $U = 2027.0 < \ucrit \approx 2078$).
We compared the average log guesses necessary for those passwords guessed
in an online attack.
We used Welch's t-test, since the variance differed between groups,
and found that this difference was statistically significant 
($\pval = 0.0044$).
However, for offline attacks, there was not a statistically significant 
difference ($\pval > 0.1$). 
All results are shown in \cref{tab:results:whitebox}.

\subsection{Memorability}
\label{sec:results:mem}
We show success rates for all password types, login times,
and exclusion criteria in \cref{fig:results:mem}.

Immediately after password creation, users who created \SysName passwords
had the lowest login success rate out of all password types
with all exclusion criteria.
This difference was statistically significant
($\pval < 0.05$ between \SysName and control passwords for all participants
and excluding those based on keystrokes, and $\pval < 0.005$
for all other comparisons with \SysName passwords)
as analyzed with post-hoc Dunn with Bonferroni p-value adjustment.
One week later, the best performing group was either
the control group, with the strict exclusion criteria, 
or \SysName group, for the other two exclusion criteria.
The difference between any groups one week later was not statistically
significant. We compare our results to those of 
related work in \cref{sec:discussion:comparison}.

\subsection{Tool Usability}
\label{sec:results:usability}
\subsubsection{Tool Satisfaction}
\label{sec:results:usability:satisfaction}
\begin{figure}[h]
    \centering
    \includegraphics[width=\linewidth]{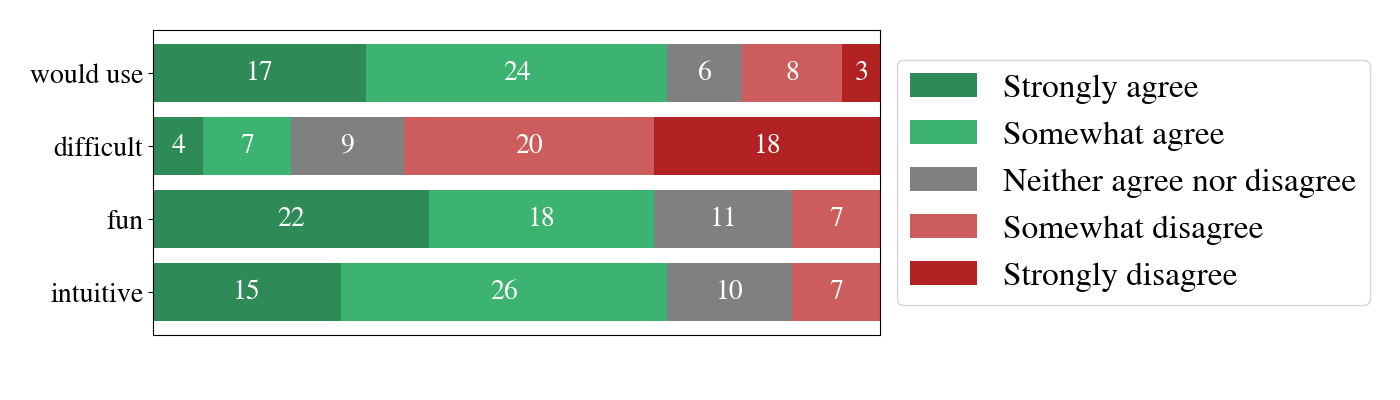}
    \caption{Survey results for tool satisfaction.
    Most participants agree with the positive statements 
    (``I would use a password created by this tool.'',
    ``The tool was fun to use.'', and
    ``The tool was intuitive to use.'')
    and disagree with the negative statements
    (``The tool was difficult to use.'').
    }
    \Description{Tool Passwords overlap less}
    \label{fig:results:satisfaction}
\end{figure}
Around 70\% of users agreed that the tool was intuitive and fun to use and said 
they would use the tool, and less than 20\% disagreed for those same questions.
18.97\% said that the tool was difficult to use, and 65.52\% disagreed.
Exact numbers are shown in \cref{fig:results:satisfaction}.
\subsubsection{Pasting Rates}
\label{sec:results:usability:pasting}
\begin{table}[!ht]
    \centering
    \begin{tabular}{@{}r@{\hspace{1.5em}}c@{\hspace{0.75em}}c@{}}
    \toprule
        \textbf{password type} & 
        \textbf{immediately after} & \textbf{one week later} \\ 
    \midrule
        \textbf{registration} & ~3.48\% & - \\
        \textbf{old} & 18.00\% & 51.22\% \\
        \textbf{control} & 16.95\% & 27.27\% \\
        \textbf{tool} & 16.99\% & 52.00\% \\
    \bottomrule
    \end{tabular}
    \caption{
    Portion of users that pasted their passwords,
    separated by password type and login event. ``Immediately after''
    occurs in Stage 1 for registration and old password, and in stage
    2 for the replacement passwords (control and tool).
    There was not a statistically significant difference at either login time
    between groups. 
    }
    \label{tab:results:paste}
\end{table}
We compared how many participants, out of those who successfully logged
in, typed their password in manually and how many 
pasted their password.
Immediately after creating the password,
about 83\% of both tool and control passwords and 82\% of old 
passwords were manually typed in,
as shown in \cref{tab:results:paste}.
A week later, however, 
about half of tool and old passwords were typed in, but 72.73\% 
of control passwords were.
More passwords were manually typed immediately 
after creating them than a week later, across password types.

We compared pasting rates across password types at the same login time:
this meant we compared old passwords at Stage 1 with tool and control
passwords at Stage 2 (``immediately after'' category), 
and we separately compared old passwords at Stage 2 with tool and control
passwords at Stage 3 (``one week later'' category).
A Kruskal-Wallis test showed no statistically significant
difference ($\pval > 0.1$) between password types
both immediately after passwords were created and a week later.
\subsubsection{Login Times}
\label{sec:results:usability:time}
\begin{table}[!ht]
    \centering
    \begin{tabular}{@{}r@{\hspace{2.25em}}r@{\hspace{0.75em}}r@{\hspace{0.75em}}r@{}}
    \toprule
        & \textbf{overall} & \textbf{typing time} 
        & \textbf{time/char} \\ 
    \midrule
        \multicolumn{4}{c}{\textbf{immediately after}} \\ 
        old & 12.92 & ~5.75 & 0.51 \\ 
        control & 11.24 & ~5.71 & 0.50 \\ 
        tool & 62.52 & 12.57 & 0.78 \\ 
    \midrule
        \multicolumn{4}{c}{\textbf{one week later}} \\ 
        old & 40.65 & 19.03 & 1.74 \\ 
        control & 25.13 & ~7.00 & 0.60 \\ 
        tool & 38.16 & 16.28 & 0.77 \\ 
    \bottomrule
    \end{tabular}
    \caption{
        Average times in seconds separated by login event and password type.
        Immediately after, login time is longest for tool passwords,
        with a statistically significant difference.
        One week later, users take longest to log in with original passwords.
    }
    \label{tab:results:time}
\end{table}
Average overall time, typing time, and time-per-character
are shown in \cref{tab:results:time}.
We found that overall time and typing time increased
a week after the passwords were created.

There was one outlier for old passwords a week later 
that took significantly longer than other cases, over 10 minutes.
Without this outlier, the overall time became 29.72s,
typing time became 7.9s, and time per character became 0.72s.

We performed Welch's ANOVA on typing time 
between old, control, and tool passwords immediately after and
one week later. We found no statistically significant difference 
one week later. Immediately after, we found a statistical effect
with $\pval=0.0006$. We performed ad-hoc analysis with Tukey's HSD
and found a statistically significant difference ($\pval < 10^{-6}$)
between both the tool and control and tool and old passwords.
This was expected, as the typing time for the tool group was much longer.

\section{Discussion}
\label{sec:discussion}
In this section, we first compare the security and memorability results 
of \SysName to past works (\cref{sec:discussion:comparison}).
Then, we describe the implications of \SysName
and how well it can apply to real-world use cases 
(\cref{sec:discussion:implications}),
as well as the ways in which our results may be limited 
(\cref{sec:discussion:limitations}).

\begin{figure*}[h]
    \centering
    \resizebox{\linewidth}{!}{\input{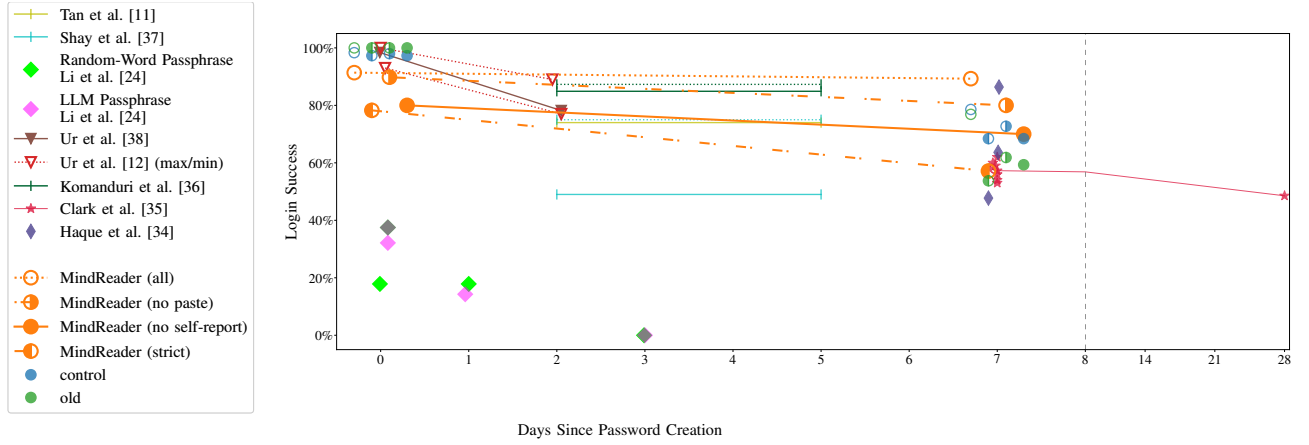}}
    \caption{Comparison of password recall
    for \SysName and previous work. Results also shown in 
    \cref{fig:appendix:memtable}. x-values are slightly adjusted for visibility.
    Our passwords are tested at the same time or later than most comparisons 
    and showed comparable recall rates at the same time. 
    }
    \label{fig:discussion:comparison:recall_comparison}
\end{figure*}

\begin{figure*}[h]
    \centering
    \resizebox{\linewidth}{!}{\input{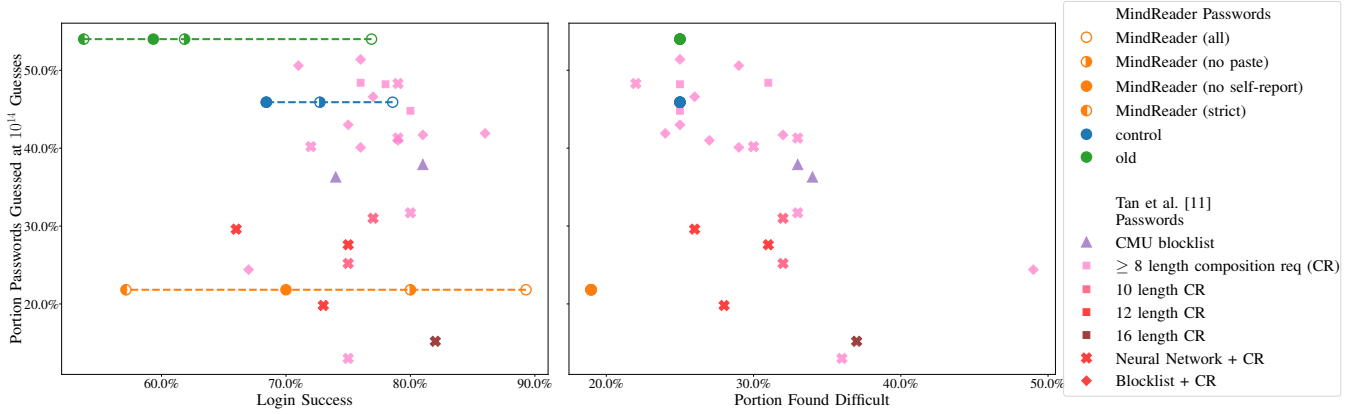}}
    \caption{
        Comparison of password recall, security, and user difficulty
        for \SysName (7-day recall circular markers) 
        and~\citet{Tan:2020:Practical} (2-5 day recall).
        For~\citet{Tan:2020:Practical} 
        triangular markers, only the 
        Carnegie Mellon blocklist was used.
        For the other markers, composition requirements
        (number of character types and minimum length)
        were used. The color of these markers corresponds to the minimum
        length. User difficulty for \SysName control and old passwords
        uses the difficulty found by~\citet{Tan:2020:Practical} for users
        who created passwords of at least length 8 and 1 character type.
        Fewer \SysName passwords were guessed than most points of comparison.
        Fewer participants found \SysName difficult to use
        than any other password type.
    }
    \label{fig:discussion:sec_recall_comparison}
\end{figure*}

\subsection{Comparison to Previous Work}
\label{sec:discussion:comparison}

We begin by comparing the memorability results of \SysName to those of related
work on passwords or passphrases. 
All points of comparison for password recall are shown 
in \cref{fig:appendix:memtable}.
Password recall is often measured at less than 7 days past password
creation~\cite{Komanduri:2011:People,Shay:2012:Horse,Ur:2012:Measure,
Ur:2017:Driven,Tan:2020:Practical,Li:2025:LLM}, 
but some other papers also use 7 days
as the period of check-in~\cite{Taiabul:2017:Bits,Clark:2025:List}
and~\citet{Clark:2025:List} is notable for measuring password recall
at 28 days past password creation. Related work generally excludes data for 
users who are measured to paste their password in or report copying it 
in~\cite{Komanduri:2011:People,Shay:2012:Horse,Ur:2012:Measure,Taiabul:2017:Bits,
Ur:2017:Driven,Tan:2020:Practical,Clark:2025:List} or asks participants
to not write the passwords down~\cite{Li:2025:LLM}. A limit of five attempts
is commonly used~\cite{Komanduri:2011:People,Shay:2012:Horse,
Ur:2012:Measure,
Ur:2017:Driven,Tan:2020:Practical,Clark:2025:List}, 
with some exceptions~\cite{Taiabul:2017:Bits,Li:2025:LLM}.

We present how \SysName passwords compare in recall and recall time
to these works in \cref{fig:discussion:comparison:recall_comparison}.
Out of papers that report memorability for all participants, without
the exclusion of users who pasted their password, \SysName had the highest
password memorability of 89.29\%, compared to 74.38-82.96\%~\cite{Shay:2012:Horse}
and 87.30\%~\cite{Komanduri:2011:People}. 
Among participants who did not paste or write down their password previously,
\SysName passwords had 57.14\% recall, 
compared to 35.00-89.00\% in related works~\cite{Shay:2012:Horse,
  Tan:2020:Practical,Li:2025:LLM} 
at $< 7$ days
and 57.28\% for the only other work
that reports memorability at 7 days~\cite{Clark:2025:List}.

\begin{table*}[t]
    \centering
    \begin{tabular}{@{}r@{\hspace{3em}}c@{\hspace{0.75em}}c@{\hspace{0.75em}}c@{\hspace{0.75em}}c@{}}
    \toprule
        & similar password & same method & new method & other \\ 
        total count & \textbf{30} & \textbf{69} & 24 & 3 \\ 
    \midrule
        I used multiple words & \textbf{6} & \textbf{17} & 9 & 0 \\ 
        I used the abbreviation of a phrase or an abbreviation of multiple words 
        & 5 & 10 & 0 & 0 \\
        I used a single word. & \textbf{6} & \textbf{8} & 4 & 1 \\
        I replaced characters with similar ones (e.g. replacing “I” with “1”) 
        & 3 & 13 & 0 & 0 \\
        I added random numbers. & 8 & 21 & 9 & 0 \\
        I used only numbers. & 1 & 0 & 0 & 0 \\
        The entire password was random characters. 
        & \textbf{4} & \textbf{14} & 5 & 2 \\ 
        I used personal information, like a name or year. 
        & \textbf{13} & \textbf{17} & 5 & 1 \\ 
        None of the above (please specify) & 3 & 9 & 2 & 1 \\ 
    \bottomrule
    \end{tabular}
    \caption{
    Responses of participants during the pre-replacement survey.
    Rows are responses to 
    ``Which of the following describes the method you 
    used to come up with your original password? Select all that apply.''
    and columns are responses to
    ``Which of the following best describes the method 
    you used to come up with your original password, 
    in comparison to other passwords you have used?''
    Respondents could choose multiple responses for the first question.
    Bolded table entries are discussed in \cref{sec:discussion:implications}.
    }
    \label{tab:discussion:implications:survey_response}
\end{table*}

Next, we compare password security to two other works that use 
PGS to measure password guessability,
those being~\citet{Ur:2017:Driven} and~\citet{Tan:2020:Practical}. 
These papers generally report guessability up to $10^{14}$ guesses.
At $10^{14}$ guesses, 21.82\% of \SysName passwords are guessed with PGS,
based on the minimum guess number between all methods.
\citet{Tan:2020:Practical} reports on 25 password types. Out of these,
three have a smaller proportion guessed at $10^{14}$ guesses. 
All of these have a neural network strength requirement---guaranteeing 
that none of them are estimated to be guessed in $10^6$, $10^{10}$, 
or $10^{12}$ guesses.
Additionally, for all three types, at least 38\% of participants
reported password creation to be annoying and at least 28\% found it difficult.
By comparison, 19\% of our participants found using \SysName difficult,
and 25\% of participants in \citet{Tan:2020:Practical} found creating
a password with the only requirement being a minimum length of eight characters
(the same as the old and manually replaced passwords in our study)
difficult.
For \citet{Ur:2017:Driven}, two password types explored out of 15 have fewer
passwords guessed in $10^{14}$. Both of these require passwords 
with three character types, 12 total characters, 
and show a strict password strength bar.
\citet{Tan:2020:Practical} also reports portion of 
passwords guessed at $10^{6}$ guesses. For two password types, no passwords 
are cracked. For the remaining types, between 0.03 and 6.3\% are guessed.
No \SysName passwords are guessed at $10^{7}$ guesses.
We show a comparison of memorability and security between 
\citet{Tan:2020:Practical} and \SysName in 
\cref{fig:discussion:sec_recall_comparison},
as well as a comparison of security and user difficulty.

\subsection{Real-World Applicability}
\label{sec:discussion:implications}

\SysName is not necessarily the best solution for all users who 
need to replace their passwords. For example, users who already
use securely generated passwords and auto-fill their passwords
would not benefit from the memorability or security of \SysName passwords.

However, we believe a large fraction of users would benefit, including
password-manager users for the passwords that they do not auto-fill
(e.g., to their password manager).
Of the participants who completed the pre-replacement survey
(see \cref{sec:appendix:user:survey:pre}),
a majority (69/126)
reported that they replaced their passwords using the same method that they 
usually use to create their original passwords.
Another 30 participants
used ``a password similar to an existing password''.
These users' results are of interest to us, since this indicates
the realism of their results when using \SysName.
Of both of those groups,
only 18 said the entire password was random 
but 23 said they used multiple words, 14 said they used a single word,
and 30 said they used personal information,
all of which result in passwords that can be made more secure by \SysName.
This suggests that there are indeed real-world users who
would benefit from \SysName.
More results from the pre-replacement survey  
are shown in \cref{tab:discussion:implications:survey_response}.

\subsection{Limitations}
\label{sec:discussion:limitations}

\paragraph{Ecological validity}
\label{sec:discussion:limitations:user}
We measured the security and memorability of \SysName passwords
through a user study. Although we followed best practices in password
research~\cite{Komanduri:2011:People},
participants in our
study had different incentives than real users and so the passwords
participants created in our study may have different characteristics
than real passwords.

We measured memorability only immediately after password creation and
a week later, and so our results may not generalize to passwords used
more often or for longer periods of time. 

Additionally, we were not able to evaluate \SysName passwords in the 
context of a user managing a set of passwords. 
While our methodology is standard practice in password research
(c.f., \cite{Shay:2012:Horse,Ur:2012:Measure,Tan:2020:Practical,Li:2025:LLM,Ur:2017:Driven,Clark:2025:List,Taiabul:2017:Bits}),
evaluations along the 
lines of Vu et al.~\cite{Vu:2007:Accounts}
would provide additional context.

\paragraph{Sample size}
\label{sec:discussion:limitations:data}
We found that \SysName passwords were statistically significantly stronger than
original passwords and control-condition replacements against a
black-box attacker and against a white-box attacker in online attacks.
However, we did not find a statistically significant difference for
offline white-box attacks, or for memorability for return logins,
pasting rates, or login times immediately after creation. A larger
sample could potentially lead to statistically significant results for
those comparisons.

Additionally, we found a statistically significant difference 
showing that \SysName passwords were slightly less memorable and took
slightly longer to type \textit{immediately} after they were created.
We believe that these results are less important when it comes to 
long-term usability of \SysName passwords than results on logins
that take place a week later; nonetheless, these results
indicate that there may be opportunity to further improve \SysName so
that it outperforms traditional password replacement even more
comprehensively.

\section{Conclusion}
\label{sec:conclusion}

We introduced \SysName, a tool and method for helping users replace
their passwords. Through a user study, we studied the passwords that
participants created both with and without \SysName.
Participants who used \SysName were generally satisfied: 
a majority said they would use it and that it was fun and intuitive;
a minority said it was difficult to use.
We found that passwords created with \SysName were more secure 
against a traditional attacker than manually replaced passwords
in both the online and the offline settings.
Moreover,
passwords created with \SysName were also more secure
than those created without it by an online attacker with knowledge of both
\textit{the original password} and \textit{\SysName's design}.
Finally, passwords created with \SysName were comparably 
memorable to those created without it when users returned to log in.
\section{Ethical Considerations}
\label{sec:ethics}

We identify three main categories of potential stakeholders impacted
by this research: user study participants (\cref{sec:ethics:user}),
victims of password leaks (\cref{sec:ethics:breach}),
and potential users of \SysName (\cref{sec:ethics:society}).

\subsection{User Study Participants}
\label{sec:ethics:user}
The main concern for harms that can affect participants is violation of privacy.
We mitigated any potential harm to participants in the following ways.
In our user study, we did not collect any direct identifiers about
our participants. The only indirect identifiers we collected---age 
and technological proficiency---were collected as they could be
relevant to this research. We did so to minimize the risk
of participants being identified.
We asked participants to not use any passwords that they use for any
real world account, whether in the past or present. 
Nonetheless, the passwords used may have been similar to a real-world password,
so we do not release the passwords used.

We also maintained respect for our participants' personhood through informed 
consent, repeated reminders and availability of information for 
the user study procedure and purpose, and allowing users to discontinue
their participation at any point. Users were paid separately for each 
stage and therefore not penalized if they did not want to participate
in later stages.

A user study was necessary to appropriately evaluate the effectiveness of 
\SysName. We believe we were able to maintain ethical principles in
our human subjects research and appropriately mitigate any harm 
to participants in a way that justifies our work on \SysName.

\subsection{Password Breach Victims}
\label{sec:ethics:breach}
While we only tested \SysName with passwords that were voluntarily provided
(either through our own informal testing or the user study)
and not used for real-world accounts, real-world passwords 
from password breaches were used in developing some of the 
security evaluation methods used in this paper.

Building on research that has used stolen passwords means that our work
benefits from the violation of these victims' privacy, even if neither our
work nor the work we built on \textit{caused} these violations. 
To mitigate any additional harm and prevent further 
privacy violation to victims, we did not publicize or mention the 
exact breaches these passwords originated from.

Ultimately, we believe that using these datasets in a research capacity 
to evaluate \SysName is justifiable. These datasets already exist and 
are available publicly, and using methods informed by these 
datasets to evaluate \SysName password security provides us more informed
and more accurate estimations for security, especially against 
attackers who also take advantage of these breaches.

\subsection{Potential \SysName Users and Society at Large}
\label{sec:ethics:society}
The final potential stakeholder affected by this research 
is users of \SysName and society at large if \SysName were to be implemented
outside of a research context.
A key risk of using \SysName is an attacker who takes 
advantage of her knowledge of the tool to crack passwords 
of \SysName users. We evaluate this risk in \cref{sec:results:whitebox}
and take care to iterate the design of \SysName to 
be resilient against this attack, mitigating the risk of privacy
violation if a \SysName password were to be guessed.
We also acknowledge that, while we found that \SysName passwords 
were more resilient to these online attacks than other passwords,
we did not find statistically significant improvement for 
offline attacks and attacks in general. We therefore emphasize the need
for further research before widespread usage of \SysName.

We also acknowledge a potential risk in the usage of \SysName---that is 
that passwords, or at the very least password segments, need to be available
in plaintext in order to find replacements. An appropriate 
deployment would address this risk. \SysName may need to be 
implemented so it can be run locally or accessed through a trusted party.
However, these details on the real-world deployment of \SysName are
outside the scope of this work.

\bibliographystyle{IEEEtranN}
\bibliography{full,refs}

\onecolumn

\section*{Appendix}

\begin{table*}[h]
    \centering
    \begin{tabular}{@{}r@{\hspace{0.5em}}r@{\hspace{0.5em}}r@{\hspace{2em}}r@{\hspace{1em}}r@{\hspace{1em}}r@{\hspace{1em}}r@{\hspace{2em}}c@{\hspace{0.5em}}c@{\hspace{0.5em}}c@{}}
    \toprule
    & & & \multicolumn{4}{c}{recall rate (\%) $\uparrow$} & \\
    password [+citation] & password type & time frame & no paste or self-report & no paste & no self-report & all & attempts \\
    \midrule
    MindReader & \multirow{3}{*}{password} & \multirow{3}{*}{7 days} & 57.14 & 80.00 & 70.00 & 89.29 & \multirow{3}{*}{5} \\
    Control & & & 68.42 & 72.73 & 68.42 & 78.57 & \\
    Original & & & 53.75 & 61.86 & 59.34 & 76.88 & \\
    \midrule
    ~\citet{Tan:2020:Practical}  & passwords & \multirow{3}{*}{2-5 days} & 66-86 & & & & ? \\
    ~\citet{Shay:2012:Horse} & passwords & & 52.00-65.00 & & & 74.38-80.94 & 5 \\
    ~\citet{Shay:2012:Horse} & passphrases & & 35.00-57.00 & & & 65.51-82.96 & 5 \\
    \midrule 
    \multirow{3}{*}{~\citet{Li:2025:LLM} (LLM)} & \multirow{3}{*}{passphrase} & 1 hour & && 32.14-37.50 && \multirow{3}{*}{1} \\
    & & 1 day & && 14.29 && \\
    & & 3 days & && 0.00 && \\
    \midrule
    \multirow{3}{*}{~\citet{Li:2025:LLM} (Random-Word)} & \multirow{3}{*}{passphrase} & 1 hour & && 17.86-37.50 && \multirow{3}{*}{1} \\
    & & 1 day & && 17.86 && \\
    & & 3 days & && 0.00 && \\
    \midrule
    ~\citet{Ur:2017:Driven} & \multirow{3}{*}{password} & 2 days & 78.20 & & & & \multirow{3}{*}{5} \\
    ~\citet{Ur:2012:Measure} & & 2 days & 77.00-89.00 & & & & \\
    ~\citet{Komanduri:2011:People} & & 2-5 days & 84.90 & & & 87.30 & \\
    \midrule 
    \multirow{2}{*}{~\citet{Clark:2025:List}} & \multirow{2}{*}{password} & 7 days & 57.28 & & & & \multirow{2}{*}{5}\\
    & & 28 days & 48.54 & & & & \\
    \midrule 
    ~\citet{Taiabul:2017:Bits} (control) & \multirow{3}{*}{password} & \multirow{3}{*}{7 days} & & & 47.73 & & \multirow{3}{*}{3} \\
    ~\citet{Taiabul:2017:Bits} (loci) & & & & & 63.64 & & \\
    ~\citet{Taiabul:2017:Bits} (link)& & & & & 86.37 & & \\
    \bottomrule
    \end{tabular}
    \caption{More detailed results from \cref{fig:discussion:comparison:recall_comparison}.}
    \label{fig:appendix:memtable}
\end{table*}

\begin{figure*}[h]
    \centering

    \begin{subfigure}{0.5\textwidth}
    \centering
    \includegraphics[scale=0.25, frame]{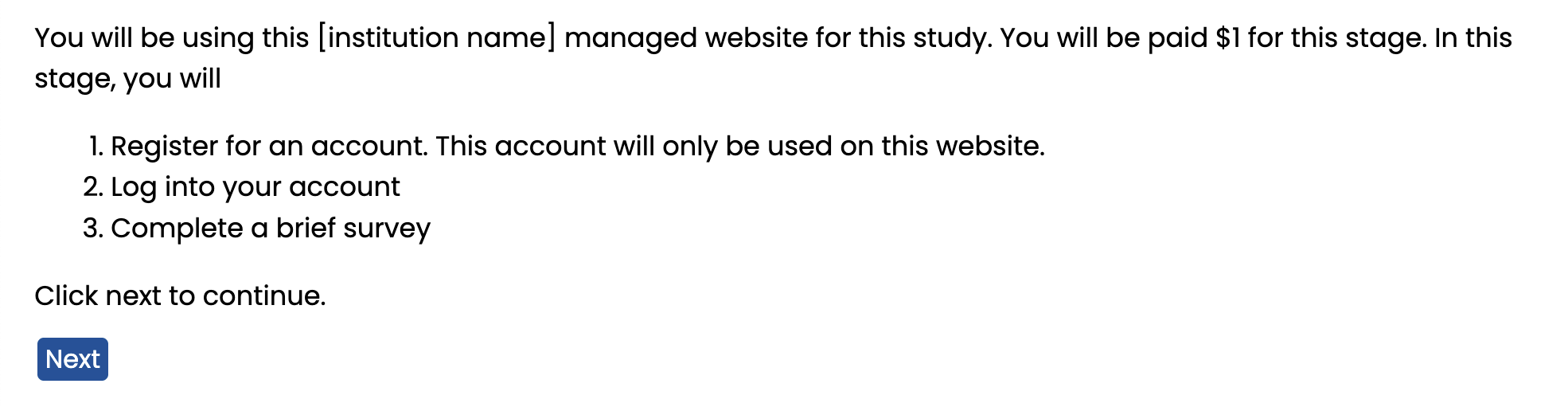}
    \caption{A summary of the current stage shown to participants before
    beginning stage 1 of the study.}
    \label{fig:appendix:tool:intro1}
    \end{subfigure}
    \hfill
    \begin{subfigure}{0.25\textwidth}
    \centering
    \includegraphics[scale=0.25, frame]{
        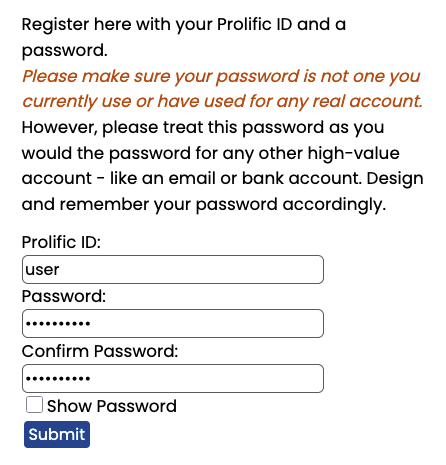}
    \caption{\SysName's registration page}
    \label{fig:appendix:tool:registration}
    \end{subfigure}
    \hfill
    \begin{subfigure}{0.2\textwidth}
    \centering
    \includegraphics[scale=0.25, frame]{
        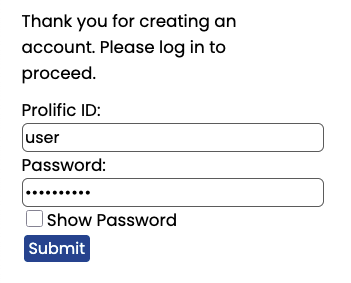}
    \caption{Login page, shown immediately after registering before
    the first login event.}
    \label{fig:appendix:tool:logins1}
    \end{subfigure}

    \caption{Samples of the \SysName UI during stage 1}
    \label{fig:appendix:tool:stage1}
\end{figure*}
\begin{figure*}[h]
    \centering

    \begin{subfigure}{0.5\textwidth}
    \centering
    \includegraphics[scale=0.25, frame]{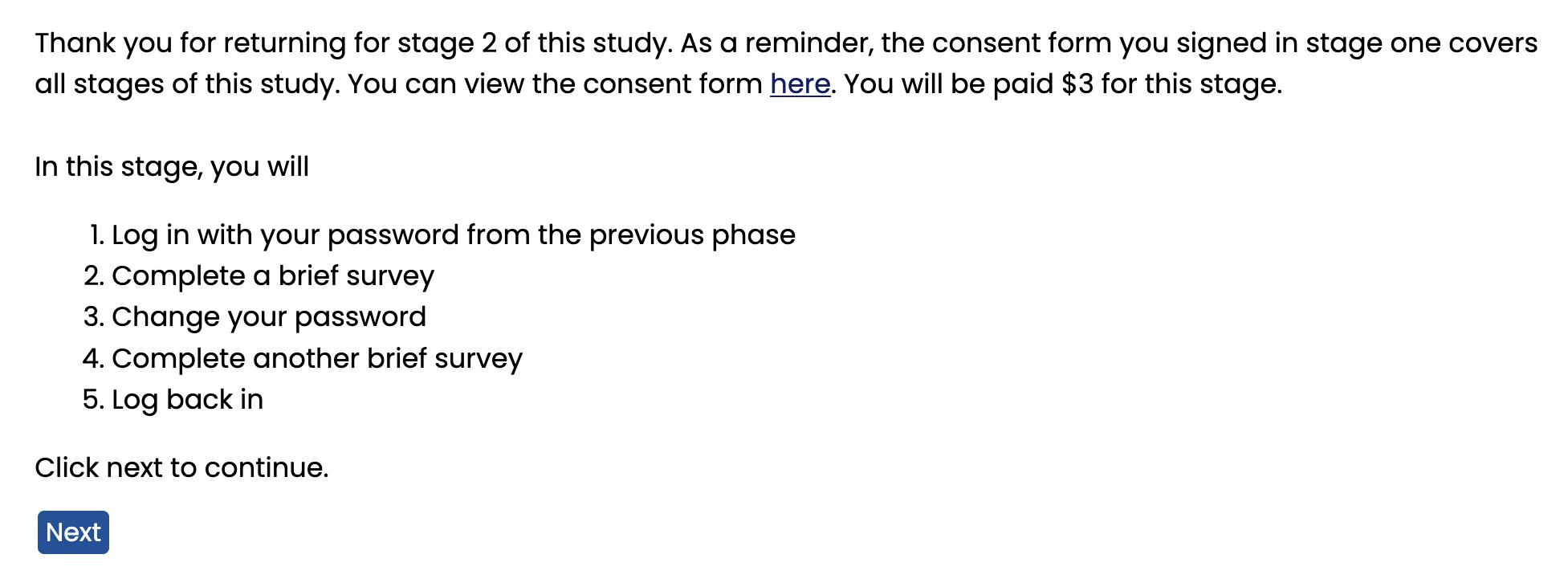}
    \caption{An introduction shown to users right before
    stage 2 of the study. This screenshot shows the view for the tool group.
    Control group participants see a different compensation value.
    A link to the consent form is available.}
    \label{fig:appendix:tool:intro2}
    \end{subfigure}
    \hfill
    \begin{subfigure}{0.2\textwidth}
    \centering
    \includegraphics[scale=0.25, frame]{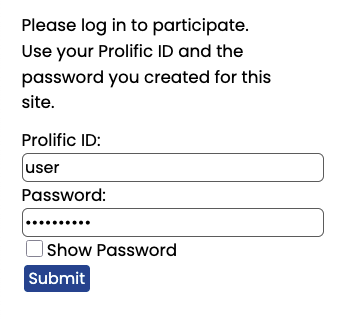}
    \caption{The login page shown to users in stage 2 of the study.}
    \label{fig:appendix:tool:login}
    \end{subfigure}
    \hfill
    \begin{subfigure}{0.25\textwidth}
    \centering
    \includegraphics[scale=0.25, frame]{
        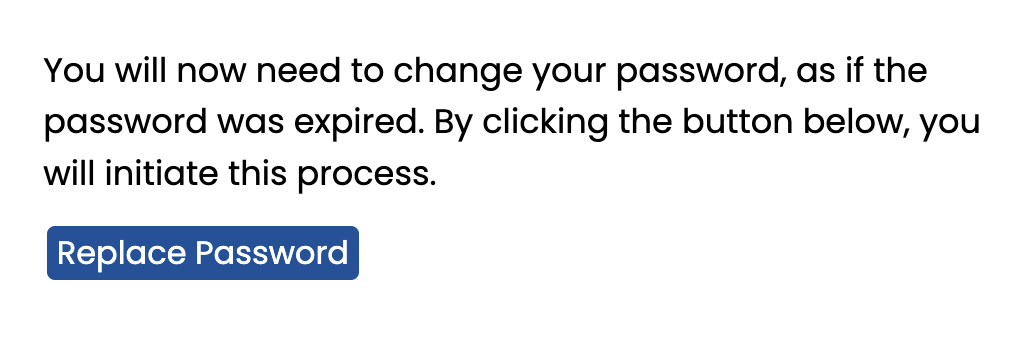}
    \caption{Information shown to users right before they replace their password.
    Participants in the tool group are shown \cref{fig:tool:segmentation} next.}
    \label{fig:appendix:tool:pre}
    \end{subfigure}

    \caption{Samples of the \SysName UI during stage 2, not shown in 
    \cref{sec:tool:approach}}
    \label{fig:appendix:tool:stage2}
\end{figure*}
\begin{figure}[h]
    \centering
    \includegraphics[scale=0.3, frame]{
        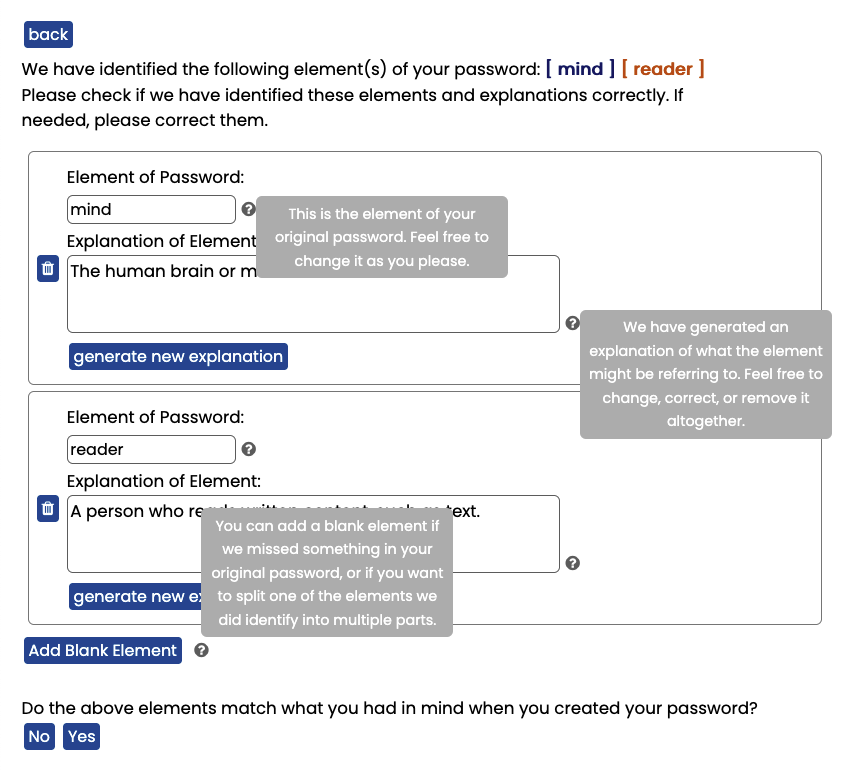}

    \caption{Tooltips from the password segmentation step.}
    \label{fig:appendix:tool:tooltips_segmentation}
\end{figure}
\begin{figure}[h]

    \centering
    % \begin{subfigure}{0.5\textwidth}
    \begin{subfigure}{0.4\textwidth}
    \centering
    \includegraphics[scale=0.3, frame]{
        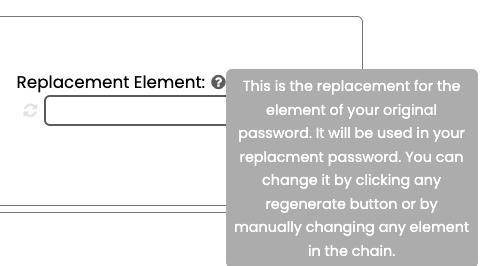}
    \caption{Tooltip next to the segment replacement step}
    \label{fig:appendix:tool:tooltips_replacement_2}
    \end{subfigure}
    \hfill
    % \begin{subfigure}{0.5\textwidth}
    \begin{subfigure}{0.4\textwidth}
    \centering
    \includegraphics[scale=0.25, frame]{
        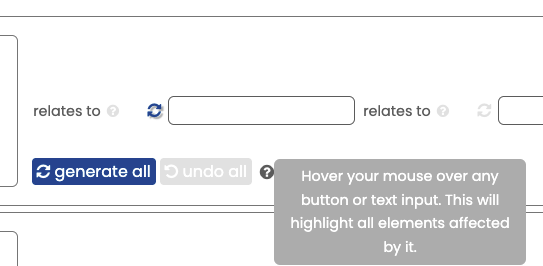}
    \caption{Tooltip next to the control buttons for each segment}
    \label{fig:appendix:tool:tooltips_replacement_1}
    \end{subfigure}
    \hfill
    % \begin{subfigure}{0.5\textwidth}
    \begin{subfigure}{0.5\textwidth}
    \centering
    \includegraphics[scale=0.3, frame]{
        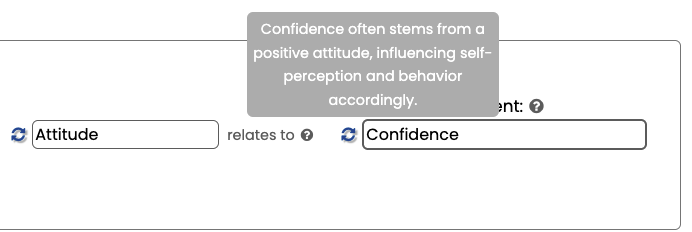}
    \caption{A tooltip explaining the connection between
    two links in the chain of association.}
    \label{fig:appendix:tool:tooltip_link}
    \end{subfigure}

    \caption{Samples of tooltips from the segment replacement step}
    \label{fig:appendix:tool:tooltips_replacement}
\end{figure}

\section{Tool Details}
\begin{table*}[t]
    \centering
    \begin{tabular}{p{6cm} p{11cm}}
    \toprule
        use-case & prompt \\
    \midrule
        confirming if leetspeak was used 
        & ``Is the passphrase `\{pwd\}' written in leetspeak? 
        Respond in one word, just `yes' or `no'.'' \\ \midrule
        removing leetspeak 
        & ``Remove the leetspeak from this password, when appropriate: 
        `\{pwd\}'. Return the password, and nothing else.'' \\ \midrule
        generating segment explanation 
        & ``What might `\{segment\}' be referring to? 
        Respond with just the answer, under 10 words.'' \\ \midrule
        default prompt for getting next element in chain
        & ``For a user that has \{elem\} in their password, 
        what is a closely related string that they can use instead?
        Return one word, and nothing else.'' \\ \midrule
        \multirow{4}{5cm}{alternative prompts for 
        getting next element in chain} 
        & ``For a user that has \{elem\} in their password, 
        what is a similar word that they can use instead?'' \\ 
        & ``What is a word or string related to \{elem\}?'' \\ 
        & ``What is the first word that you think of when you hear \{elem\}?'' \\
        & + ``Return one word, and nothing else.'' (added to all prompts
        in this category) \\ \midrule
        generating connections between two chain elements 
        & ``What is a simple but truthful and fact-based connection between 
        \{item1\} and \{item2\}? Respond with at most 15 words.'' \\
    \bottomrule
    \end{tabular}
    \caption{Prompts used by \SysName. \{elem\} is ``\{segment\}'' if 
    the segment has no explanation and 
    ` ``{segment}'' ({explanation})' otherwise. All alternative prompts
    for getting next element in the chain end with 
    ``Return one word, and nothing else.''. These alternative prompts
    are used if there were multiple unsuccessful attempts in generating
    a unique chain element.
    }
    \label{tab:appendix:tool:prompts}
\end{table*}
\subsection{Tool UI}
\label{sec:appendix:tool:ui}
In this section, we provide views of the \SysName UI not shown in 
\cref{sec:tool:approach}.
These are separated by stage.
Views from stage 1 are shown in \cref{fig:appendix:tool:stage1}
and feature the introduction page, registration, and login pages.
Views from stage 2 are shown in \cref{fig:appendix:tool:stage2},
and include the introduction, login, and pre-replacement page
that informs users they will need to replace their passwords.

We also show example tooltips, which are visible while hovered over,
with additional information for the users in 
\cref{fig:appendix:tool:tooltips_segmentation} and 
\cref{fig:appendix:tool:tooltips_replacement}.

\subsection{Prompts}
\label{sec:appendix:tool:prompts}
We specifically provide prompts used by \SysName in 
\cref{tab:appendix:tool:prompts}.

\newpage
\section{User Study}
\label{sec:appendix:user}
In this section, we provide all user study material used 
(\cref{sec:appendix:user:survey})
and provide compiled information for the demographic data collected 
(\cref{sec:appendix:user:demographics}).

\subsection{User Study Survey Questions}
\label{sec:appendix:user:survey}
\subsubsection{Demographics Replacement Survey}
\label{sec:appendix:user:survey:demo}
This survey was taken by participants during stage 1 after they 
completed the consent form, registered for an account,
and logged in with their newly created password.
After this survey, participants were done with stage 1.
\begin{enumerate}
    \item Select what applies 
    in relation to your experience with password managers:
    \begin{itemize}
        \item I am not familiar with password managers.
        \item I am familiar with password managers, but I do not use them.
        \item I am familiar with password managers, and I use them.
        \item Other (please specify): (text entry)
    \end{itemize}

    \item How frequently do you give computer or technology advice 
    (e.g., to friends, family, or colleagues)?
    \begin{itemize}
        \item Daily
        \item A few times a week
        \item A few times a month
        \item Rarely
        \item Never
    \end{itemize}

    \item Which of the following best describes 
    your educational background or job field?
    \begin{itemize}
        \item I have an education in, 
        or work in, the field of computer science, engineering or IT
        \item I do not have an education in, 
        or work in, the field of computer science, engineering or IT
    \end{itemize}

    \item Select your age group:
    \begin{itemize}
        \item 18-24
        \item 25-34
        \item 35-44
        \item 45-54
        \item 55-64
        \item 65+
    \end{itemize}
\end{enumerate}

\subsubsection{Pre-Password Replacement Survey}
\label{sec:appendix:user:survey:pre}
This survey was taken by participants during stage 2 after they 
logged in with their password from the previous week (original password).
After this survey, participants replaced their password.
\begin{enumerate}
    \item Did you write down your password in order to remember it? 
    This will not affect your compensation.
        \begin{itemize}
            \item No
            \item Yes
        \end{itemize}

    \item Did you copy your password in? 
    This will not affect your compensation.
        \begin{itemize}
            \item Yes---I typed it in manually from a reference
            \item Yes---I used copy/paste or autocomplete
            \item No---I typed it in from memory
        \end{itemize}

    \item Which of the following best describes the method 
    you used to come up with your original password, 
    in comparison to other passwords you have used?
        \begin{itemize}
            \item I used a password similar to an existing password.
            \item I used a completely new password, 
            but used a similar method to what I normally do.
            \item I used a completely different password, 
            and I used a method I do not normally use.
            \item Other (please specify): (text entry)
        \end{itemize}

    \item Which of the following describes the method you 
    used to come up with your original password? Select all that apply.
        \begin{itemize}
            \item I used the abbreviation of a phrase/multiple words.
            \item I used a combination of a phrase or multiple words.
            \item I used a single word.
            \item I replaced characters with similar ones 
            (e.g. replacing “I” with “1”)
            \item I added random numbers.
            \item I used only numbers.
            \item The entire password was random characters.
            \item I used personal information, like a name or year.
            \item None of the above (please specify): (text entry)
        \end{itemize}

    \item Feel free to provide any additional 
    information on how you came up with your original password. (text entry)
\end{enumerate}

\subsubsection{Pre-Password Replacement Survey (Control)}
\label{sec:appendix:user:survey:control}
This survey was taken by participants in the control group
during stage 2 after they replaced their password.
After this survey, participants were directed to 
log in with their new password.
\begin{enumerate}
    \item Which of the following most closely describes your method for replacing your password?
        \begin{itemize}
            \item I used my old password, and made minimal changes.
            \item I came up with my new password using my old password, 
            but the passwords themselves are distinct.
            \item My new password is not related to my old password, 
            but I used the same logic or method to come up with both.
            \item My new password is not related to my old password, 
            and I used different logic or method to come up with them.
            \item Other (please specify): (text entry)
        \end{itemize}

    \item Feel free to provide any additional 
    information on how you came up with your replacement password. (text entry)
\end{enumerate}

\subsubsection{Pre-Password Replacement Survey (Tool)}
\label{sec:appendix:user:survey:tool}
This survey was taken by participants in the tool group
during stage 2 after they replaced their password with \SysName.
After this survey, participants were directed to 
log in with their new password.
\begin{enumerate}
    \item The tool was intuitive to use. 
        (strongly agree to strongly disagree, 5 point scale)
    \item The tool was fun to use. 
        (strongly agree to strongly disagree, 5 point scale)
    \item The tool was difficult to use. 
        (strongly agree to strongly disagree, 5 point scale)
    \item I would use a password created by this tool. 
        (strongly agree to strongly disagree, 5 point scale)
    \item Please describe any difficulties you had with this tool. 
        (text entry box)
    \item Feel free to provide any more information on how you came up 
    with your replacement password using the tool. (text entry)

\end{enumerate}

\subsubsection{Recall Survey}
\label{sec:appendix:user:survey:recall}
This survey was taken by participants in the control group
during stage 3 after they logged back in with their replacement password.
After this survey, participants were done with the study.
\begin{enumerate}
    \item Did you write down your new password in order to remember it? 
    This will not affect your compensation.
        \begin{itemize}
            \item No
            \item Yes
        \end{itemize}

    \item Did you copy your new password in? 
    This will not affect your compensation.
        \begin{itemize}
            \item Yes---I typed it in manually from a reference
            \item Yes---I used copy/paste or autocomplete
            \item No---I typed it in from memory
        \end{itemize}
\end{enumerate}

\subsection{User Study Demographics}
\label{sec:appendix:user:demographics}

\begin{table}[!ht]
    \centering
    \begin{tabular}{@{}r@{\hspace{1.5em}}c@{\hspace{0.5em}}c@{\hspace{0.5em}}c@{}}
    \toprule 
        Question & Response & Count \\
    \midrule
        Are you familiar & Yes, and I use them. & 131 \\
        with password & Yes, but I do not use them. & 54 \\
        managers? & No. & 15 \\
    \midrule
        How frequently  & Rarely & 53 \\
        do you give computer & A few times a month & 68 \\
        or technology advice & Daily & 23 \\
        (e.g., to friends, & A few times a week & 54 \\
        family, or colleagues)? & Never & 2 \\
    \midrule
        Do you have an education  & No & 121 \\
        in, or work in, the field & & \\
        of computer science, & Yes & 79 \\
        engineering or IT? & & \\
    \midrule
        \multirow{6}{*}{Select your age group}  & 25-34 & 78 \\
        & 18-24 & 31 \\
        & 45-54 & 26 \\
        & 35-44 & 47 \\
        & 55-64 & 13 \\
        & 65+ & 5 \\
    \bottomrule
    \end{tabular}
    \caption{Responses to the demographics survey.}
    \label{tab:appendix:user:demographics}
\end{table}
We provide user responses to demographic questions from 
\cref{sec:appendix:user:survey:demo}
in \cref{tab:appendix:user:demographics}.
Participants varied in age but were most likely
to be 25-34 years old, mostly did not have a background 
in computer science, engineering, or IT, and most reported that
they use password managers.

\end{document}